\documentclass[aps,prb,twocolumn,superscriptaddress,amsfonts,amssymb,showpacs]{revtex4-1}

\usepackage{mhchem}
\usepackage{graphicx}
\usepackage{dcolumn}
\usepackage{amssymb}
\usepackage{wasysym}
\usepackage{color}
\usepackage{chemformula}
\usepackage{tabularx} 
\usepackage{longtable}
\usepackage{booktabs} 
\usepackage{siunitx}
\usepackage{booktabs} 

\begin{document}


\title{Copper ion dynamics and phase segregation in Cu-rich tetrahedrite: an NMR study}
\keywords{Solid State NMR, Ionic Conductivity, Thermoelectric Materials, Structure Transformation}



\author{Nader Ghassemi}
\affiliation{Department of Physics and Astronomy, Texas A\&M University, College
Station, TX 77843, USA}
\author{ Yefan Tian}
\affiliation{Department of Physics and Astronomy, Texas A\&M University, College
Station, TX 77843, USA}
\author{Xu Lu}
\affiliation{Department of Applied Physics, Chongqing University, Chongqing, 401331, China}
\author{ Yanci Yan}
\affiliation{Department of Applied Physics, Chongqing University, Chongqing, 401331, China}
\author{ Xiaoyuan Zhou}
\affiliation{Department of Applied Physics, Chongqing University, Chongqing, 401331, China}
\author{Joseph H. Ross, Jr.}
\affiliation{Department of Physics and Astronomy, Texas A\&M University, College Station, TX 77843, USA}
\altaffiliation{Materials Science and Engineering Department,  Texas A\&M University, College Station, TX 77843, USA}
\email{ross@physics.tamu.edu}


\date{\today}

\begin{abstract}
$^{63}$Cu NMR measurements are reported for the Cu-rich phase of \ch{Cu_{12+x}Sb4S13} ($x  \lesssim 2$) and compared to \ch{Cu12Sb4S13}. We identify the NMR signatures of the phase segregation into Cu-poor ($x \approx 0$) and Cu-rich ($x \lesssim 2$) phases, with the metal-insulator transition observed in \ch{Cu12Sb4S13} suppressed in the Cu-rich phase. Based on NMR $T_1$ and $T_2$ measurements, the results demonstrate Cu-ion hopping below room temperature with an activation energy of $\sim$150 meV for the Cu-rich phase, consistent with superionic behavior. The NMR results also demonstrate the effects of Cu-ion mobility in the \ch{Cu12Sb4S13} phase, but with a larger activation barrier. We identify a small difference in NMR Knight shift for the metallic phase of \ch{Cu12Sb4S13}, compared to the Cu-rich phase, and when compared to DFT calculations the results indicate a mix of hyperfine contributions to the metallic shift.
\end{abstract}

\maketitle

\section{\label{sec:level1} Introduction}
The necessity for new clean energy sources is becoming more and more apparent due to depletion of fossil resources and environmental degradation. Thermoelectric materials provide the simplest basis for collecting energy from wasted heat and reducing greenhouse emissions, however most of the current thermoelectric materials are made of toxic, expensive and/or scarce elements. Improved thermoelectric compounds should also have high Seebeck coefficients and electric conductivity, combined with low thermal conductivity to achieve the needed efficiencies. Recently tetrahedrite, a natural mineral based on the \ch{Cu12Sb4S13} compound, identified as a potentially significant new material, with high thermoelectric performance at midrange temperatures \cite{suekuni2013high,lu2013high, PremKumar2017, SUN2019835}.
This material includes nontoxic, earth-abundant elements and can also be doped by transitional metals.

The phase diagram of \ch{Cu_{12+x}Sb4S13} with $0\le x\le2$ has been studied over a wide range of  temperatures. Samples segregate into two tetrahedrite structural phases (\ch{Cu12Sb4S13} and \ch{Cu14Sb4S13} ) which converge to a single phase \cite{vaqueiro2017influence, tatsuka1977tetrahedrite, johnson1983brillouin} at higher temperatures, driven by relatively mobile Cu ions. In \ch{Cu12Sb4S13}, the Fermi level sits within the valence band making it metallic or heavily p-type, and a metal semiconductor phase transition (MST) occurs \cite{Suekuni_2012} at around 85 K, with structural changes accompanied by an increase in resistivity \cite{Suekuni_2012} as well as a drop in magnetic susceptibility \cite{DiBenedetto2005}. The extra coppers in \ch{Cu14Sb4S13} are expected to fill unoccupied states and push the Fermi level to the band-gap, hence rendering the compound insulating. Recently, \citeauthor{YAN2018127} reported that adding 1.5 extra Cu per formula unit to \ch{Cu_{12}Sb4S13} enhances the efficiency up to $\sim$ 66$\%$. In further band engineering, it was found that adding Se resulted in $\sim$ 64 $\%$ further enhancement in power factor \cite{YAN2018127}.
A new method was also proposed in tetrahedrites to lower $\kappa_L$ by using spinodal decomposition \cite{bouyrie2015exsolution} of the Cu-poor and Cu-rich phases. 
This induces additional channels for phonon scattering and reduces the lattice thermal conductivity substantially. In addition, there are ongoing efforts to utilize Cu ion mobility in these and related Cu chalcogenides as an avenue for reducing thermal conductivity for thermoelectric applications \cite{Jana2018}, as well as in other device applications \cite{Munjal_2019}, although the impact of mobility on the stability of microfabricated devices is also a potential issue. Thus it is important to understand the kinetics and local structures involved in this segregation process.

NMR can be an effective way to probe ionic hopping in solids \cite{brinkmann1992nmr, hull2004superionics}, allowing sensitivity to a range of dynamical time-scales from $10^{-7}$ to $1$ s and longer, not easily probed by other techniques. The local chemical information provided through NMR studies also provides a useful complement to other techniques such as diffraction studies, and also can often allow disordered or mixed systems to be studied effectively \cite{tang1999diffusion}. Nuclei with quadrupole moments, such as the \ce{^{63}Cu} nucleus probed here, can be particularly sensitive in this regard, since the electric field gradients which couple to the quadrupole moment can exhibit large changes in response to atomic displacements. In this work, we have used NMR techniques to study a copper rich \ch{Cu_{12+x}Sb4S13} tetrahedrite material. We show that the NMR spectra and relaxation times demonstrate the presence of the Cu motion extending to low temperature, with particularly high mobility observed in the copper rich tetrahedrite. Spin-lattice relation measurements provide an estimation of the activation energy of these mobile ions.

\section{\label{sec:level2} Experiment}
\subsection{\label{sec:level2-1} Sample preparation}
Synthesis: polycrystalline samples of \ch{Cu12Sb4S13} (Cu poor) and \ch{Cu_{13.5}Sb_{3.98}Sn_{0.02}S13} (Cu rich) tetrahedrite were obtained by melting stoichiometric amount of high purity ($>99.99\%$) elements (Cu, Sb, S) at $923$ K for $12$ h, and then cooling down to room temperature. The obtained samples were annealed at $723$ K for a week. Finally, the obtained ingots were hand ground into fine powder for spark plasma sintering (SPS-625) at $673$ K for $5$ min under a uniaxial pressure of $45$ MPa.

\begin{table*}[t]
  \centering
  \begin{tabular}{|c c c c c c c@ {\hspace{0.4 cm}} c c c c|}  \toprule
  & & &\multicolumn{2}{c}{Cu12 phase} & & & &  \multicolumn{2}{r}{Cu14 phase} &  \\ \cline{3-6}\cline{8-11}
Site &Wyckoff position&x  & y & z & SOF&  & x  & y & z & SOF \\ \toprule
Cu-I& 12d&0.25 & 0.5 & 0 & 1& & 0.25& 0.5 & 0 &  0.9873  \\
Cu-II & 12e &  &  &  & & & 0.2142 & 0& 0 & 0.9460 \\
Cu-II & 24g & -0.2485  & 0.0642 & -0.0642 & 0.5 & &  &  &   & \\
Cu-III & 24g&  &  &  &  &    & 0.2851 &0.2851 &0.0102 & 0.15\\
Sb & 8c & 0.2652 & 0.2652 & 0.2652 & 1 & & 0.2666 & 0.2666 & 0.2666 & 0.9881\\ 
Sn & 8c &  &  &  &  & & 0.2653 & 0.2653 & 0.2653 & 0.005\\ 
S-I  & 24g & 0.1193  & 0.1193  & 0.3788  & 1 & & 0.1144 & 0.1144 & 0.3513 & 1\\
S-II & 2a & 0    & 0  & 0 &1& & 0    & 0  & 0 &1\\ \toprule
  \end{tabular}
  \caption{Crystallographic information at 300 K for Cu rich sample including Cu12 minority phase. Lattice parameters $a = \SI{10.4409}{\angstrom}$  for Cu14 (phase fraction 70$\%$) and $a = \SI{10.3205}{\angstrom}$ for Cu12 (phase fraction 30 $\%$). $R_w=6.05 \%$. SOF:  site occupancy factor. }
  \label{tab: X_ray}
\end{table*}

\subsection{\label{sec:level2-2} Measurement methods}

 Room temperature powder X-ray diffraction (XRD) data were collected using a Bruker D8 spectrometer with CuK$\alpha$ radiation and analyzed using the GSAS-II package\cite{toby2013gsas}. NMR measurements were carried out using spin echo integration versus frequency and by assembling and superposing Fast Fourier Transformation (FFT) spectra at a sequence of frequencies. These measurements were executed in a magnetic field of $8.9$ T at frequencies near $100$ MHz, using a custom-made pulse spectrometer in a temperature range from $4.2$ to $300$ K. A silver coil was used to eliminate spurious Cu NMR signals.
\ce{^{63}Cu} NMR chemical shifts were referenced to \ch{CuCl}.

NMR spin-lattice relaxation time ($T_1$) measurements were obtained based on an inversion recovery sequence implementing a multi-exponential function for recovery of the
central transition \cite{19761}. We follow the convention $K = ( f - f_0)/ f_0$ for Knight shifts, with $f_0$ the standard reference frequency and positive shifts having paramagnetic sign. In the analysis, we used nuclear moment values ($Q$ and $\gamma$) reported in
Ref. \cite{harris2002nmr}. Electric field gradients (EFGs) are given in terms of the standard parameters $\nu_Q= \frac{3eQV_{zz}}{2I(2I-1)h}$ and $\eta=(V_{xx}-V_{yy})/V_{zz}$.

\section{ {Results and Analysis}}
\begin{figure}[t]
\includegraphics[width=8cm]{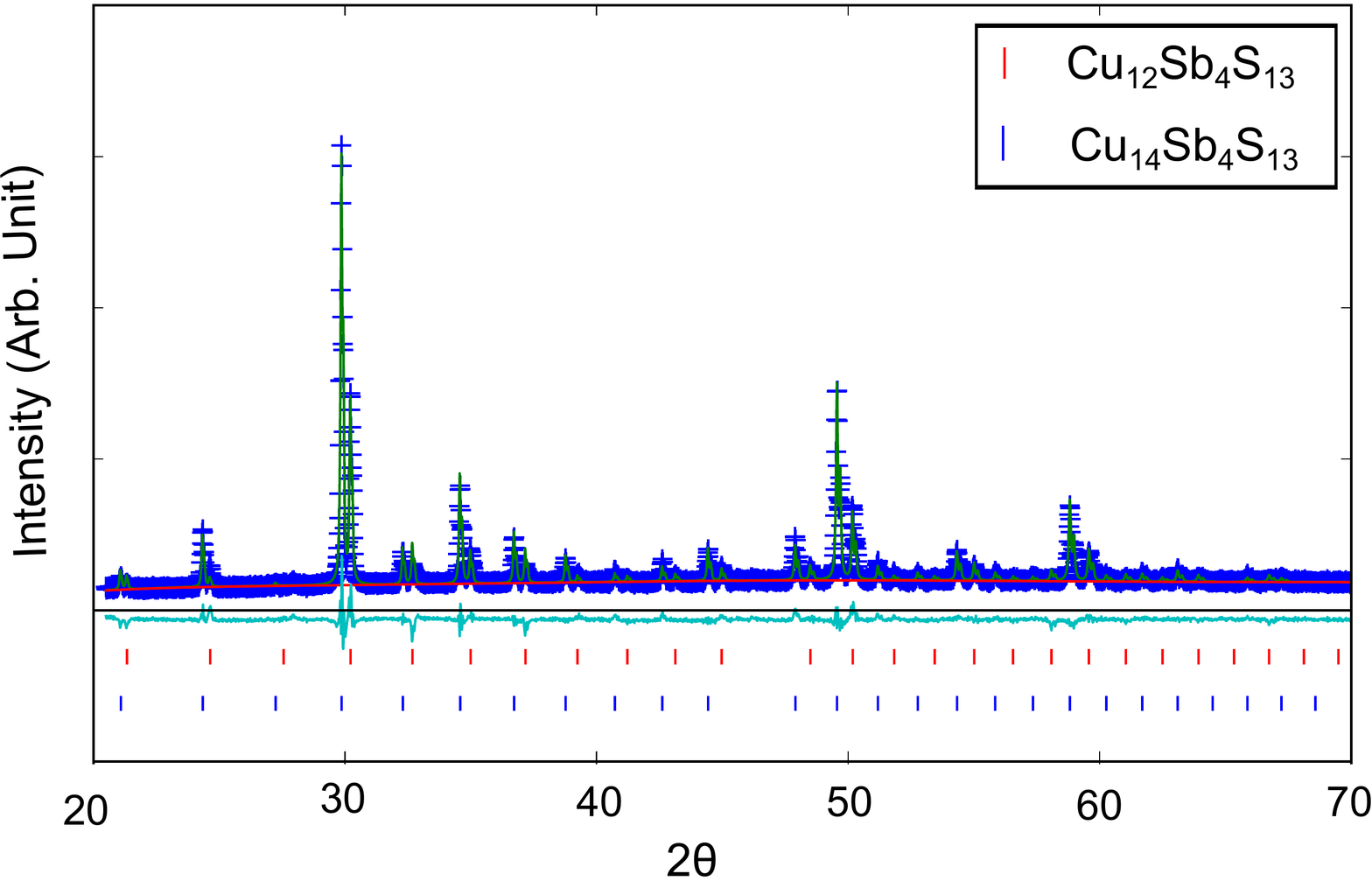}
\centering
\caption{ Powder X-ray diffraction patterns collected at room temperature for the \ch{Cu_{13.6}Sb_{3.98}Sn_{0.02}S13} sample, showing the two fitted phases. }
\label{fig: x_ray}
\end{figure}

The X-ray result and refinement are plotted in Figure \ref{fig: x_ray}.
XRD analysis shows that the sample includes distinct \ch{Cu_{13.6}Sb4S13} and \ch{Cu12Sb4S13} phases without any secondary phases. Detailed results are in Table \ref{tab: X_ray}. These phases will be referred to as Cu14 and Cu12 respectively. Both phases were refined in the cubic structure of space group $I\overline{4}m$ (217) with lattice constants $a=\SI{10.4409}{\angstrom}$ and $a=\SI{10.3205}{\angstrom}$ respectively, in good agreement with what was previously reported  \cite{vaqueiro2017influence, ghassemi2018structure}. The Cu14 majority phase has phase fraction $0.70$ and the Cu12 minority phase $0.30$. 
The Cu12 unit cell has two distinct Cu sites while Cu14 phase has three Cu sites. 
A good fit for the Cu12 minority phase was obtained by setting all sites to $100\%$ occupancy, except for Cu-II which was modeled as having a split-site 24g configuration. 

The half-occupied 24g site-II represents a two-fold off-center position for this ion, consistent with the fit described by \citeauthor{vaqueiro2017influence}  \cite{vaqueiro2017influence} Cu14 has a similar configuration with the addition of an interstitial Cu-III partially occupied site. Note that the XRD refinement in the Cu14 phase had a low sensitivity to Cu-I occupation in the range of 0.90 to 0.99. However, we obtained a slightly better fit with $0.99$ occupation (Rw= $6.053 \%$ vs Rw=  $6.138 \%$ for $0.92$ occupation).

\begin{figure}[t]
\includegraphics[width=9 cm]{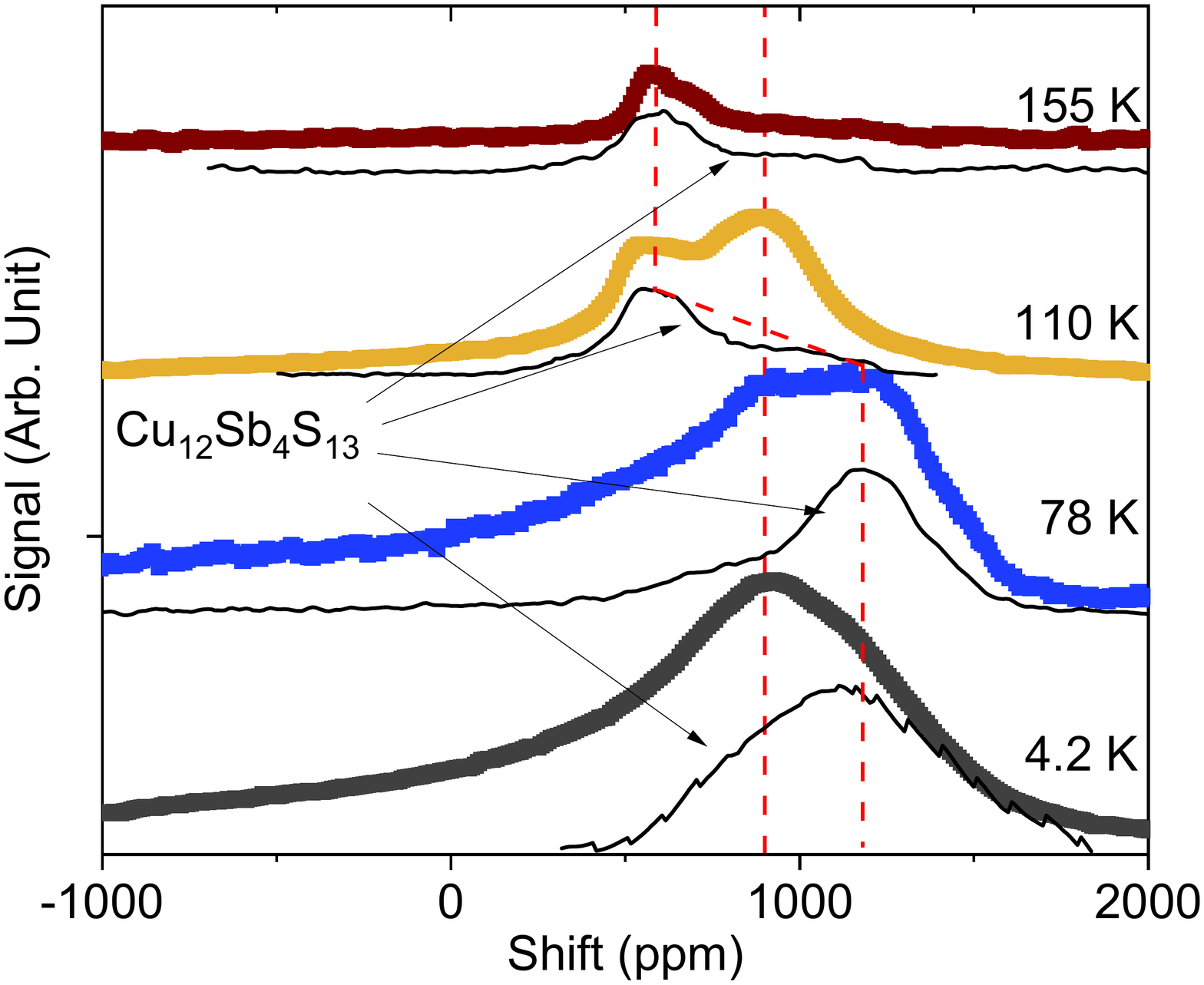}
\centering
\caption{  $^{63}$Cu NMR spectra for \ch{Cu14Sb4S13} at temperatures as shown. Solid curves are lineshapes for \ch{Cu12Sb4S13} from reference\cite{ghassemi2018structure}. The dashed lines are guides to the eye identifying the Cu-I site in Cu14 and Cu12 phases. }
\label{fig: LS_vs_T}
\end{figure}

$^{63}$Cu NMR spectra obtained at several temperatures for the \ch{Cu14Sb4S13} sample are shown in Figure \ref{fig: LS_vs_T}. 
Also superposed are spectra for a \ch{Cu12SbS13} sample, as reported previously \cite{ghassemi2018structure}. The Cu-II position in Cu12 has a large quadrupole broadening \cite{ghassemi2018structure} ($\nu_Q \approx 18$ MHz), and is well out of range in these spectra, whereas Cu-I has a very small quadrupole broadening ($\nu_Q \le 1$ MHz) \cite{bastow2015121,matsui2019first, kitagawa2015suppression}. Since Cu14 has essentially the same structure, with the addition of the interstitial site, we assume here that Cu-II for the Cu14 phase is also out of range here, and we analyze these spectra as representing Cu-I and Cu-III sites for the Cu14 phase as well as the Cu-I site for the Cu12 phase.
Above the Cu12 MST we find that the spectrum contains two main peaks. These are the central transitions ($-1/2 \rightleftharpoons 1/2$) for copper nuclei with $I = 3/2$. A broad peak is also observed to underlie these lines. Therefore, the spectra were fitted assuming three distinct sites. 

In a previous NMR study \cite{ghassemi2018structure} of \ch{Cu12Sb4S13}, two peaks with $\nu_Q \approx 3.85 $ MHz and $\nu_Q \approx 7.59$ MHz identified at room temperature were assigned to Cu-II and Cu-I respectively. However, a DFT calculation for the \ch{Cu3SbS4} compound (see supplement) shows that Cu has $\nu_Q \approx 3.64 $ MHz and $\nu_Q \approx 6.59$ MHz and we found a reflection due to small amount of this phase in the previous XRD results \cite{ghassemi2018structure} which was not identified before. Therefore, we conclude the presence of \ch{Cu3SbS4} in the previous sample explains the presence of the small peaks in room temperature. 
This phase is not found in the Cu14 sample, and also we tested a new Cu12 sample, with no room temperature signal which is in agreement with 
 \citeauthor{matsui2019first}
 \cite{matsui2019first} and {\citeauthor{kitagawa2015suppression}}
 \cite{kitagawa2015suppression}. Note that this does not affect the observed low-T rattling result \cite{ghassemi2018structure} for the dominant Cu12 phase. In the present results, the spectral weights are significantly reduced above 150 K (Figure \ref{fig: LS_vs_T}), both in the Cu12 and Cu14 phases, an effect of ionic hopping at these temperatures, as described below.

\begin{figure}[t]
\includegraphics[width=9cm]{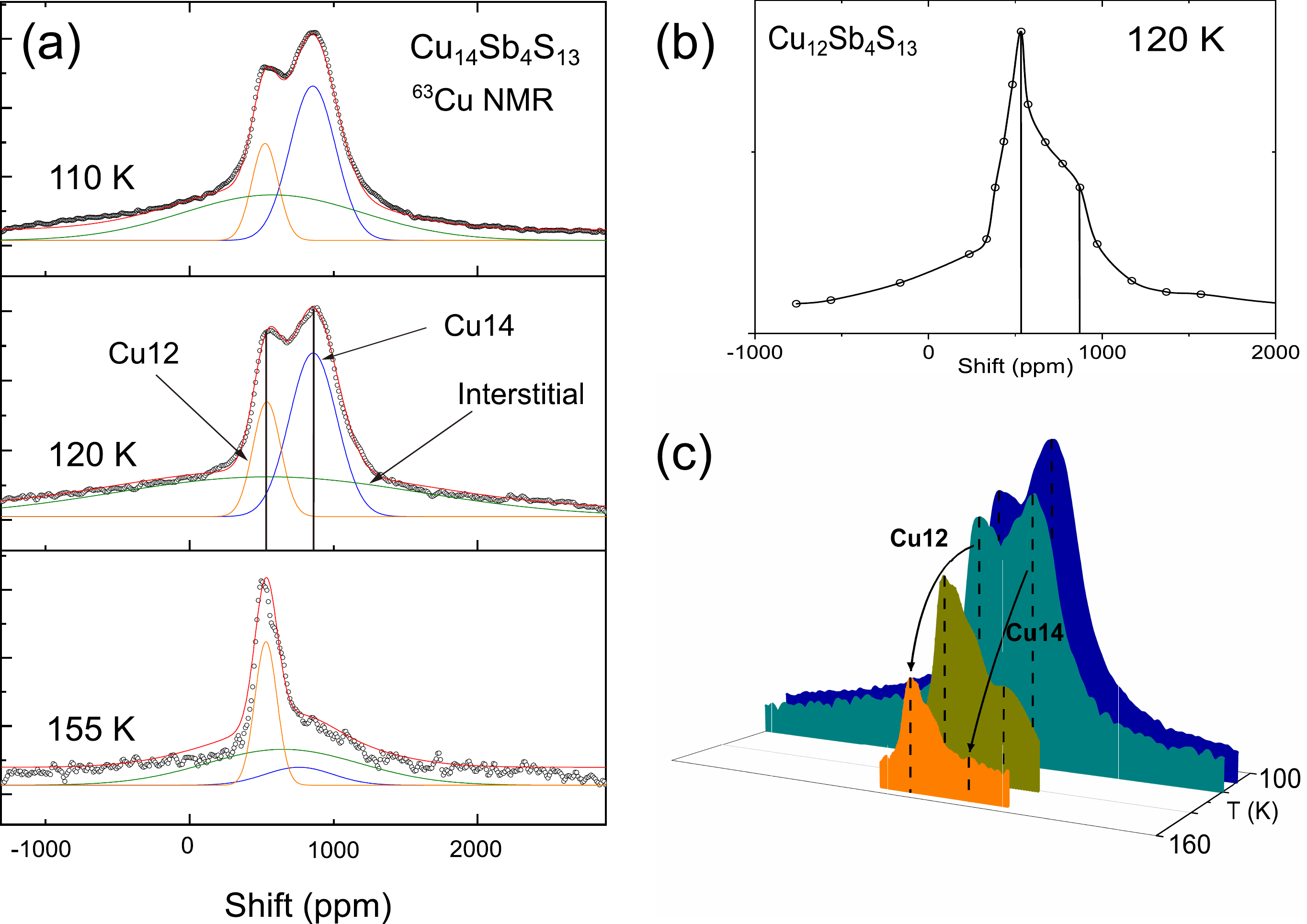}
\centering
\caption{  (a) NMR lineshapes and fitted spectra for Cu-rich sample at three indicated temperatures. Three fitted peaks are also shown, according to the model described in the text, as shown (b) Lineshape for \ch{Cu12Sb4S13} sample at 120 K with the vertical lines showing identical shift positions as those in part (a). (c) $^{63}$Cu spectra (NMR signal $\times T$) versus temperature for \ch{Cu14Sb4S13} sample showing the signal loss at high $T$. }
\label{fig: LS_Above_Tc}
\end{figure}

\begin{figure}[t]
\includegraphics[width=8cm]{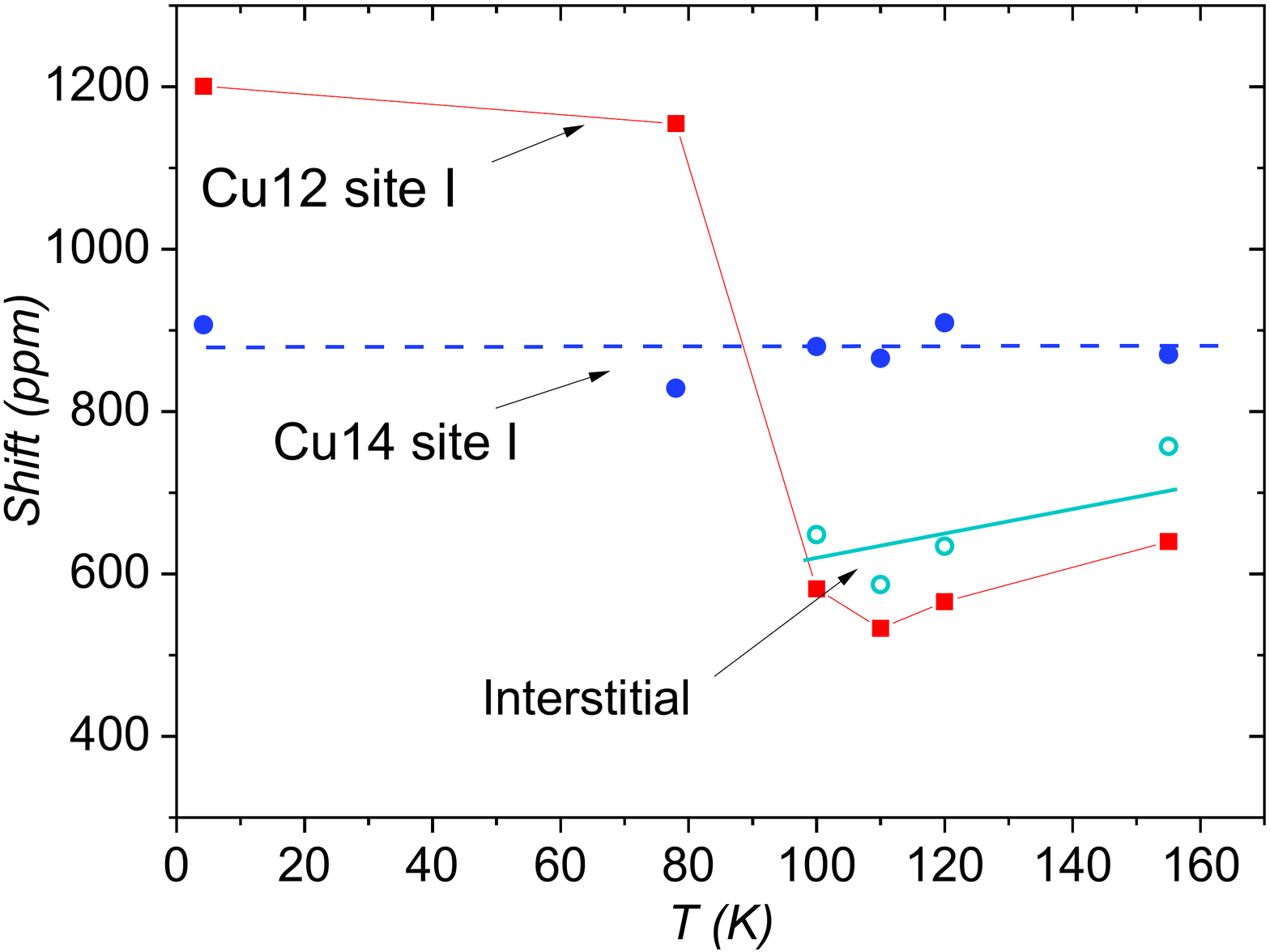}
\centering
\caption{  Temperature dependent peak positions for three fitted sites described in the text.} 
\label{fig: CS_vs_T}
\end{figure}

In the analysis of the Cu-rich sample, we fitted three Gaussian peaks to the spectra corresponding to the model identified above.  Figure \ref{fig: LS_Above_Tc}(a) shows the lineshapes with fitted peaks at three different temperatures. One of these peaks corresponds to the Cu12 minority phase, as can be seen from the superposed spectra shown in Figure \ref{fig: LS_vs_T} and also Figure \ref{fig: LS_Above_Tc}(b). The position for this peak rises with decreasing temperature from about 600 to 1200 ppm, in good agreement with the observed shifts for this composition \cite{ghassemi2018structure, matsui2019first, kitagawa2015suppression}. The remaining signal is assigned to the Cu14 main phase. At 4 K the broadening is such \cite{ghassemi2018structure} that the three peak model does not converge since the Cu12 phase also exhibits broadening due to structure change in the insulating phase. However, the general features, a shoulder at around 1200 ppm due to Cu12 and an additional peak near 900 ppm, can still be identified. These are plotted along with the fitted results in Figure \ref{fig: CS_vs_T} which summarizes the evolution of the peak positions vs temperature.

 As the temperature changes, the fitted peak near 900 ppm remains unchanged. We attribute this peak to the Cu-I site in the Cu14 majority phase; with Cu-I in the metallic phase of the Cu12 composition known to exhibit a negative Knight shift above the MST due to the core polarization mechanism, it is reasonable that this peak should have a more positive shift in the charge balanced Cu14 phase. Note also that the line-width for this peak, as well as the Cu-I peak for Cu12, are comparable to what was recently reported \cite{matsui2019first} for Cu12. As anticipated, the results indicate that the phase transition is absent in the Cu14 phase. Meanwhile, the intensity of the much broader third fitted peak is such that it is assigned to Cu-III interstitial ions combined with Cu-I sites, as described below. 
 
At 100 K, the fitted broad peak represents a fraction 0.40 of the NMR spectral weight, versus 0.36 for the Cu-I  peak in the majority phase and 0.24 for Cu-I in Cu12 phase . XRD indicates 6 Cu-I sites fully occupied in the Cu12 phase (30$\%$ phase fraction),
whereas for the Cu-rich phase the results indicate 1.8 Cu-III ions per formula unit, in addition to the 6 Cu ions on Cu-I sites. This yields an expected relative spectral weight of 0.25 for the Cu12 phase, in reasonable agreement with the fitting. By contrast, the Cu14 phase XRD occupancies yield expected relative spectral weights for Cu-I and Cu-III equal to 0.58 and 0.18 respectively. 
The corresponding fitted lines do not have this expected approximately 3:1 intensity ratio. However, the Cu-III interstitial site is a nearest-neighbor for Cu-I, and occupation of such a neighbor site would be expected to induce a large change in the EFG's experienced by the Cu-I site. With the Cu-III 24g site occupation observed to be 0.15 and since each Cu-I has four Cu-III near-neighbors, assuming random occupation of the Cu-III sites, the probability for a Cu-I to have no Cu-III neighbors will be 0.85$^4$ = 0.52, and therefore half of the Cu-I sites are expected to experience a significantly enhanced EFG. The observed reduced NMR intensity for the Cu14 phase Cu-I line, and enhancement of the broad peak intensity, are in reasonable agreement with this scenario. Thus we assume that the broad peak is due to Cu-III interstitials combined with Cu-I sites directly affected by the Cu-III occupation, with the breadth due to random occupation of the latter sites.

\section{ Copper Motion}
The temperature dependence of the \ch{Cu14Sb4S13} spectra is further illustrated in Figure \ref{fig: LS_Above_Tc}(c). Whereas the integrated spectral areas scaled by temperature would normally be temperature independent if the spectral weight is conserved, the observed reduction in NMR signal can be attributed to slow Cu hopping. This is observed in all samples: the NMR echo signal is sharply reduced in the temperature range 100 to 160 K, with the onset of hopping observed in both Cu-poor Cu12 as well as Cu-rich Cu14 in this same temperature range, although for Cu14 the faster drop implies greater Cu-ion mobility in the Cu14 phase. This can be compared to the large difference in Cu mobility reported for Cu12 vs. Cu14 phases at high temperatures once super-ionic behavior sets in \cite{vaqueiro2017influence}.

\begin{figure}[t]
\includegraphics[width=8cm]{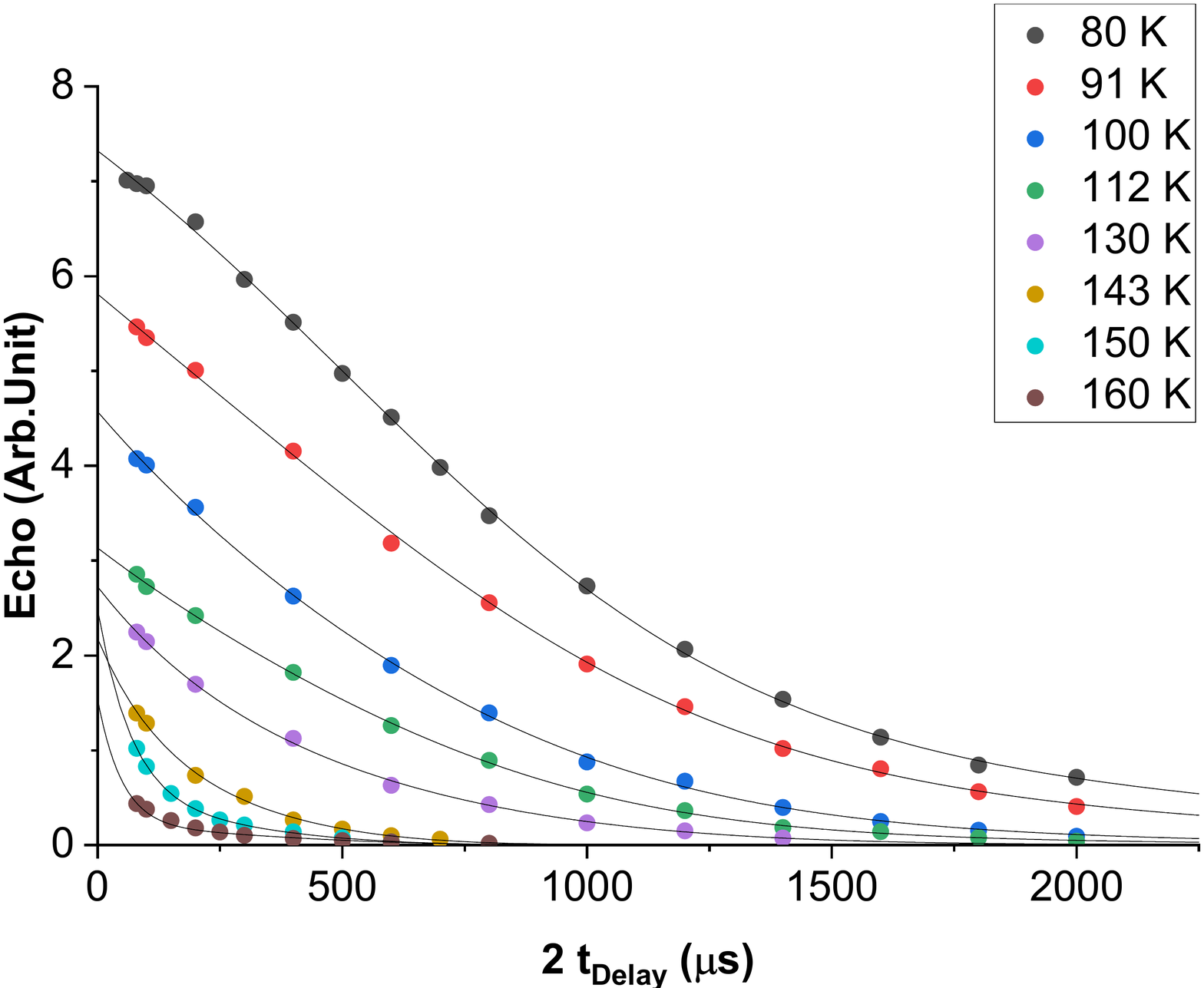}
\centering
\caption{ Spin echo decay rate for Cu14 phase at different temperatures, with fits described in text.}
\label{fig: Echo_Tdelay}
\end{figure}

For further confirmation of the dynamics of copper hopping, the spin-echo decay was measured by varying pulse separation, ($t_{del}$) in a standard spin-echo sequence, at the 900 ppm peak position for Cu-I in the Cu-rich Cu14 phase (Figure \ref{fig: Echo_Tdelay}). We also made a similar measurements for the second, Cu12-dominated sample. 
The data were fitted to 

 \begin{eqnarray}\label{eq:T2 fitting}
Echo=C[(1-\alpha) e^{-(2t_{del}/T_{2g})^2} +\alpha e^{(-2t_{del}/T_{2e})} ],
\label{eq: Guass_vs_EXP}
\end{eqnarray} 
 where $T_{2g}$ and $T_{2e}$ are the Gaussian and exponential $T_2$ decays, respectively with $T_2$ here referring generally to all processes contributing to the echo decay. The ratio $\alpha$ helps to determine the relative significance of motion, since normally exponential decay is dominant when motion is important, while more nearly Gaussian decay occurs for a static NMR line, controlled by the nuclear dipole-dipole or pseudo-dipolar couplings \cite{Abragam}.

\begin{figure}[t]
\includegraphics[width=8cm]{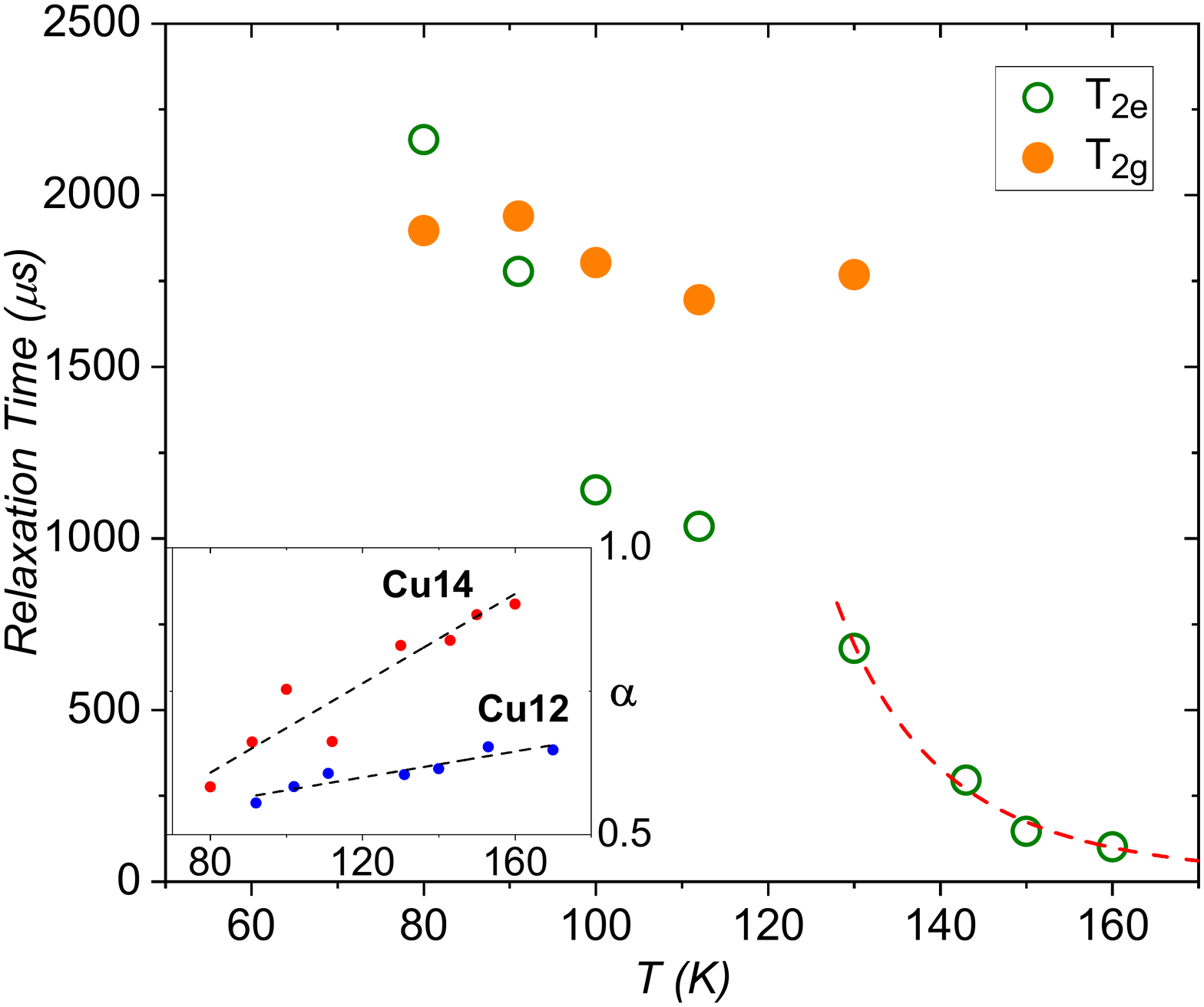}
\centering
\caption{ Fitting results of spin echo measurements. Main plot: $T_{2e}$ and  $T_{2g}$ versus T for Cu14.  The curve line is the fit described in the text. Inset: Fitting parameter $\alpha$ for majority phase in Cu rich and Cu poor samples, providing a measure of ion dynamics.}
\label{fig: Echo_Tdelay_fitting}
\end{figure}

Figure \ref{fig: Echo_Tdelay_fitting} summarizes the fitting results. The low-$T$ decay curve for both samples is dominated by Gaussian decay. However, as the temperatures rise, $\alpha$ increases, indicating that atomic motion is more important. Both samples show an increasing $\alpha$ vs. $T$, signaling the onset of dynamics, see inset of Figure \ref{fig: Echo_Tdelay_fitting}.
Note that there is a larger underlying dipole width (shorter $T_{2g}$) for the Cu12 sample contributing to the observed difference in $\alpha$ in addition to the faster increase in $1/T_{2e}$ for Cu 14. However, clearly Cu12 has a lower mobility as the majority phase, similar to what found as the minority phase in Cu14.

In analyzing the fitted $T_{2e}$ for the Cu14 main phase, we assume an activated process with hopping time $\tau= \tau_0 exp(\Delta E/k_BT)$, where $ \nu_0= 1/\tau_0$ is the attempt frequency and $\Delta E$ is the activation energy. Assuming each Cu-ion hop destroys the echo refocusing process for a Cu ion and its neighbors, $T_{2e}$ will equal the mean hopping time. This is a strong-collision, slow motion approximation, likely valid for the present case since the change of quadruple parameters associated with a sudden hopping event will make large changes in the precession frequency, removing it from the echo signal. We fitted only the last few points where $\alpha$ is close to 1, so that $T_{2e}$ can be considered as dominated by motional processes. This yielded $\Delta E \approx 116$ meV. We also obtained a relatively large attempt time $\tau_0 \approx 2.25 \times 10^{-8}$s. 
The attempt time is significantly larger than expected for vibrational motions in solids, however this is a typical situation for NMR fitting for superionic conductors in the hopping regime, for reasons which are not entirely clear \cite{BOYCE1979507, brinkmann1992nmr}. This activation energy is consistent with the result obtained from $T_1$ measurements, described below. This helps to confirm the connection of kinetic processes to the disappearance of the NMR signal, rather than for example broadening or splitting of the NMR lines due to a symmetry breaking.

\begin{figure}[t]
\includegraphics[width=8cm]{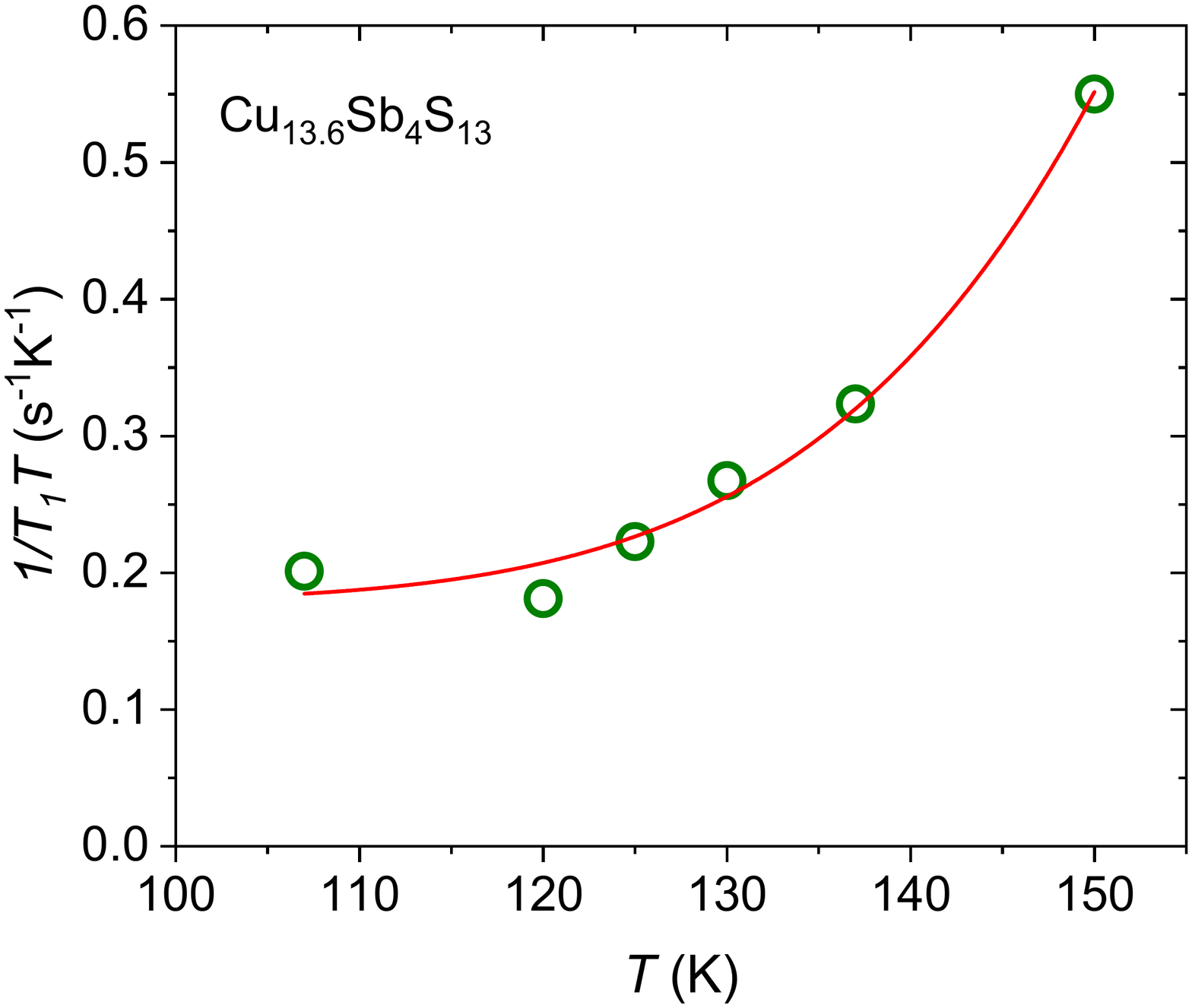}
\centering
\caption{  \ce{^{63}Cu} $1/T_1T$ for Cu-I site in Cu14 vs temperature. The curve is the activated fit explained in the text.}
\label{fig: T1_exp_fit}
\end{figure}

The nuclear dynamics can also be detected through the temperature dependence of the spin lattice relaxation time ($T_1$). \ce{^{63}Cu} $1/T_1T$ for the Cu14, Cu-I site, is shown in Figure \ref{fig: T1_exp_fit}. Faster relaxation behavior was reported in  \citeauthor{matsui2019first} \cite{matsui2019first} for a Cu12 sample although this is due in part to the difference in fitting functions. $1/T_1T$ is fitted to the activation equation \cite{BOYCE1979507,eguchi1981high} $a/T e^{(\frac{-\Delta E}{K_BT})} +b$, where $\Delta E$ is activation energy and $a$ and $b$ are constants. Figure \ref{fig: T1_exp_fit} shows results of fitting with $\Delta E = $145 $\pm$ 30 meV due to hopping in the Cu14 sample which is also in agreement with the activation energy we derived from $T_2$.  Thus even though the temperature range is limited due to the disappearance of the spin echo due to hopping, for Cu14 we obtain a consistent measure of the activated process involved here.

\section{Discussion}

 The proposed Cu14 structure \cite{makovicky1979studies}, with two extra Cu on the interstitial sites refined from X-ray diffraction, provides a good model for the observed NMR amplitudes. In the original X-ray fitting report, it had also been proposed \cite{makovicky1979studies} that some Cu ions were on unknown sites not accounted for by the Cu-I, Cu-II and Cu-III sites. Here we see that the 3-site model does well in accounting for the NMR results. Previous results \cite{vaqueiro2017influence} also point to Cu-I occupation close to 1, and as noted above, NMR is consistent with this picture.

It has also been proposed \cite{Bullett_1986} that the occupation of Cu-I should be reduced to 0.67 in the Cu14 phase with a corresponding fraction of Cu ions promoted to interstitial sites. This is due to the small refined Cu-I-Cu-III distance \cite{makovicky1979studies} of 2.37 \AA. We tested this occupation in the XRD refinement, however, the $R_w$ increased to 7.16 \% with GOF= 2.38. In the refined results of the table \ref{tab: X_ray}, bond length between Cu-I to Cu-III is 2.36 \AA, which is a normal Cu-Cu bond length \cite{chen1993metal, mehrotra1978copper}, however, with considerable uncertainty. Thus, it is unclear whether the Cu-I$-$Cu-III distance is an anomalously short one.

In the NMR spectra, we observed two peaks at $\sim$ 600  and $\sim$ 900 ppm for temperatures above the MST. Normally the contribution to the conduction electrons and holes to the NMR shift is a Knight shift ($K$). In an effective mass approximation, often appropriate
for semiconductors \cite{SIRUSI2017137}, it is found that in the metallic limit $K \approx n^{1/3}$, where 
$n$ is the carrier density. Thus, Cu14 (with balanced charge composition) is expected to have $K$ much closer to zero, and thus we expect that $K$ for the metallic phase of Cu12 is approximately equal to $-300$ ppm.

 DFT calculations (provided in Supplementary Information) for \ch{Cu12Sb4S13} show that the Cu-d partial density of states for Cu-I is $g_d(E_F)=2.48$ eV/atom. This is comparable to what was found by \textit{\citeauthor{lu2015increasing} \cite{lu2015increasing}} Considering the d-electron hyperfine field \cite{PhysRevB.54.7561} $-17.2$ T/$\mu_B$ obtained for cuprate superconductors, the calculated $g_d(E_F)$ yields an estimate $K = -2470$ ppm. The observed value is considerably smaller, thus we conclude $K$ cannot simply be due to d-core polarization based on the calculated valence band of the DOS. \citeauthor{matsui2019first} \cite{matsui2019first} have discussed that $K$ can include considerable correlation effects. However, it is surprising that such effects should give the factor of $\sim$1/6 reduction in $K$ that we obtain here.  We speculate that there is an additional competing spin contribution to the Cu12 shift due to other mechanisms. DFT results indicate that $g_s(E_F)$ is 200 times smaller than the d term, thus it seems likely that an orbital hyperfine contribution gives partial cancellation.

The activation energy for ionic motion in the Cu14 phase is comparable to that of similar types of superionic materials. Normally, solids that have ionic conductivity of 10 $S/m$ or larger as well as activation energy on order of 100 meV, are considered as solid state ionics \cite{brinkmann1992nmr}. \Citeauthor{wang2015design} indicate several examples of Li diffusion with $E_a$ in the range of $170$ to $220$ meV\cite{wang2015design}. Among Cu compounds, the activation energy in \ch{Cu2Se} is \cite{doi:10.1021/acs.jpcc.5b06079} 230 meV. In \ch{Cu2Te} the activation energy is 350 meV and doping small amount of Ag increases activation energy \cite{sirusi2016band} up to 500 meV. In CuI the extracted \cite{yesinowski2010electrical} activation energy is 640 meV but CuI has a first-order phase transition to its superionic phase. Therefore, even though tetrahedrite has a first order transformation to its superionic phase, its Cu ions are nevertheless also quite mobile at low temperature. Thus, the results are in line with those of other materials, and the NMR activation results demonstrate that the behavior extends to low temperatures, with considerable Cu mobility apparent in both phases at room temperature and below.

Based on the activation energies we consider a simplified statistical approach to the hopping process in Cu14. Since the Cu-I site is an immediate neighbor to the partially filled Cu-III sites, we assume hopping proceeds mainly through these two sites. We further assume that the Cu-I site is in the lower energy state and that the energy difference is equal to the observed hopping activation energy. The entropy is $S=k_B\hspace{0.05cm} ln\Omega$, where $\Omega$ is the multiplicity of states.
If $n_i$ is the number of Cu atoms on Cu-III interstitial sites per cell at $T=0$ and $x$ is the number of Cu-I atoms promoted to Cu-III site, we obtain:

 \begin{eqnarray}
 \Omega& = \left(\frac{(12N)!}{[(12-x) N]![x N]!}\right) \left(\frac{(24N)!}{[(24-n_i-x) N]![(n_i+x) N]!}\right)
\label{eq: stat}
\end{eqnarray} 
for a large number $N$ of unit cells. Minimizing the free energy leads to

 \begin{eqnarray}
\frac{x(n_i+x)}{(12-x)(24-x-n_i)}& \approx e^{-\Delta /KT },
\end{eqnarray}
where $\Delta$ is the energy difference between Cu-I and Cu-III sites. Using $\Delta=145$ meV derived from $T_1$ fitting results at room temperature, with $n_i= 4$ (corresponding to the \ch{Cu14Sb4S13} composition) this leads to $x=0.14$. This corresponds to Cu-I occupation of $1-(0.14/12)= 0.99$. At 493 K, the Cu-I occupation based on the same model is reduced to 0.90. The XRD results for Cu-I in the Cu14 phase shows occupation of 0.99 at room temperature which agrees with this estimation. Also \citeauthor{vaqueiro2017influence} \cite{vaqueiro2017influence} show that the occupation in their sample drops from 0.92 at room temperature to 0.86 at 493 K, before at higher T the phases become mixed in a first-order transition. The agreement seems reasonable.

For Cu12 phase the composition corresponds to $n_i=0$. Assuming an activation energy of 150 meV gives a room temperature Cu-I occupation of 0.88 which is much smaller than the observed 1.00 occupation. This agrees with our observation that the Cu12 NMR line amplitude drops more slowly vs. $T$, indicating a larger activation energy. 
Assuming instead an activation energy of 280 meV for Cu12 leads to $x=0.06$ at room temperature (thus Cu-I occupation equal to 0.99), and $x=0.54$ at 493 K (occupation of 0.95). This agrees with our XRD refinement for Cu-I in the Cu12 phase and also the reported occupation dropping to 0.95 for this phase \cite{vaqueiro2017influence} at 493 K.
Thus, the fitted activation energy results, along with an estimated larger energy for Cu12, are consistent with the reported site occupations, giving further confirmation to the analysis of the NMR results. For the Cu12 phase as noted above, the activation energy is larger by a factor on order of two, however this result is still in the range observed for other superionic conductors, and a significant amount of Cu-ion mobility is to be expected at high temperatures in both of these phases. This result should be important for device design and development.

\section{Conclusions}

In a set of Cu NMR measurements on \ch{Cu_{12+x}Sb4S13} tetrahedrites, we identified the NMR signatures of the phase segregation into Cu-poor ($x \approx 0$) and Cu-rich ($x \lesssim 2$) phases. The temperature-independent lineshape for the Cu-rich phase indicates the suppression of the phase transition for this phase. Also by observation of the NMR line shapes, combined with analysis of $T_2$ and $T_1$ relaxation behavior, we obtain a measure of Cu-ion dynamics in these phases at low temperatures. We find that the Cu ions are particularly mobile in the Cu-rich phase, with a fitted activation energy of 145 meV for ion hopping. The $x \approx 0$ phase exhibits a larger barrier for ionic motion, however in both phases we find that there is a significant rate of ionic motion at temperatures below room temperature.


\section*{acknowledgement}

This work was supported by the Robert A. Welch Foundation, Grant
No. A-1526 and National Natural Science Foundation of China (Grant Nos. 51772035, 11674040).


\bibliographystyle{apsrev}
\bibliography{arXiv.bib}

\begin{thebibliography}{42}
\expandafter\ifx\csname natexlab\endcsname\relax\def\natexlab#1{#1}\fi
\expandafter\ifx\csname bibnamefont\endcsname\relax
  \def\bibnamefont#1{#1}\fi
\expandafter\ifx\csname bibfnamefont\endcsname\relax
  \def\bibfnamefont#1{#1}\fi
\expandafter\ifx\csname citenamefont\endcsname\relax
  \def\citenamefont#1{#1}\fi
\expandafter\ifx\csname url\endcsname\relax
  \def\url#1{\texttt{#1}}\fi
\expandafter\ifx\csname urlprefix\endcsname\relax\def\urlprefix{URL }\fi
\providecommand{\bibinfo}[2]{#2}
\providecommand{\eprint}[2][]{\url{#2}}

\bibitem[{\citenamefont{Suekuni et~al.}(2013)\citenamefont{Suekuni, Tsuruta,
  Kunii, Nishiate, Nishibori, Maki, Ohta, Yamamoto, and
  Koyano}}]{suekuni2013high}
\bibinfo{author}{\bibfnamefont{K.}~\bibnamefont{Suekuni}},
  \bibinfo{author}{\bibfnamefont{K.}~\bibnamefont{Tsuruta}},
  \bibinfo{author}{\bibfnamefont{M.}~\bibnamefont{Kunii}},
  \bibinfo{author}{\bibfnamefont{H.}~\bibnamefont{Nishiate}},
  \bibinfo{author}{\bibfnamefont{E.}~\bibnamefont{Nishibori}},
  \bibinfo{author}{\bibfnamefont{S.}~\bibnamefont{Maki}},
  \bibinfo{author}{\bibfnamefont{M.}~\bibnamefont{Ohta}},
  \bibinfo{author}{\bibfnamefont{A.}~\bibnamefont{Yamamoto}}, \bibnamefont{and}
  \bibinfo{author}{\bibfnamefont{M.}~\bibnamefont{Koyano}},
  \bibinfo{journal}{Journal of Applied Physics} \textbf{\bibinfo{volume}{113}},
  \bibinfo{pages}{043712} (\bibinfo{year}{2013}).

\bibitem[{\citenamefont{Lu et~al.}(2013)\citenamefont{Lu, Morelli, Xia, Zhou,
  Ozolins, Chi, Zhou, and Uher}}]{lu2013high}
\bibinfo{author}{\bibfnamefont{X.}~\bibnamefont{Lu}},
  \bibinfo{author}{\bibfnamefont{D.~T.} \bibnamefont{Morelli}},
  \bibinfo{author}{\bibfnamefont{Y.}~\bibnamefont{Xia}},
  \bibinfo{author}{\bibfnamefont{F.}~\bibnamefont{Zhou}},
  \bibinfo{author}{\bibfnamefont{V.}~\bibnamefont{Ozolins}},
  \bibinfo{author}{\bibfnamefont{H.}~\bibnamefont{Chi}},
  \bibinfo{author}{\bibfnamefont{X.}~\bibnamefont{Zhou}}, \bibnamefont{and}
  \bibinfo{author}{\bibfnamefont{C.}~\bibnamefont{Uher}},
  \bibinfo{journal}{Advanced Energy Materials} \textbf{\bibinfo{volume}{3}},
  \bibinfo{pages}{342} (\bibinfo{year}{2013}).

\bibitem[{\citenamefont{Prem~Kumar et~al.}(2017)\citenamefont{Prem~Kumar,
  Chetty, Femi, Chattopadhyay, Malar, and Mallik}}]{PremKumar2017}
\bibinfo{author}{\bibfnamefont{D.~S.} \bibnamefont{Prem~Kumar}},
  \bibinfo{author}{\bibfnamefont{R.}~\bibnamefont{Chetty}},
  \bibinfo{author}{\bibfnamefont{O.~E.} \bibnamefont{Femi}},
  \bibinfo{author}{\bibfnamefont{K.}~\bibnamefont{Chattopadhyay}},
  \bibinfo{author}{\bibfnamefont{P.}~\bibnamefont{Malar}}, \bibnamefont{and}
  \bibinfo{author}{\bibfnamefont{R.~C.} \bibnamefont{Mallik}},
  \bibinfo{journal}{Journal of Electronic Materials}
  \textbf{\bibinfo{volume}{46}}, \bibinfo{pages}{2616} (\bibinfo{year}{2017}).

\bibitem[{\citenamefont{Sun et~al.}(2019)\citenamefont{Sun, Dong, Tang, Shang,
  Zhuang, Hu, Wu, Pan, and Li}}]{SUN2019835}
\bibinfo{author}{\bibfnamefont{F.-H.} \bibnamefont{Sun}},
  \bibinfo{author}{\bibfnamefont{J.}~\bibnamefont{Dong}},
  \bibinfo{author}{\bibfnamefont{H.}~\bibnamefont{Tang}},
  \bibinfo{author}{\bibfnamefont{P.-P.} \bibnamefont{Shang}},
  \bibinfo{author}{\bibfnamefont{H.-L.} \bibnamefont{Zhuang}},
  \bibinfo{author}{\bibfnamefont{H.}~\bibnamefont{Hu}},
  \bibinfo{author}{\bibfnamefont{C.-F.} \bibnamefont{Wu}},
  \bibinfo{author}{\bibfnamefont{Y.}~\bibnamefont{Pan}}, \bibnamefont{and}
  \bibinfo{author}{\bibfnamefont{J.-F.} \bibnamefont{Li}},
  \bibinfo{journal}{Nano Energy} \textbf{\bibinfo{volume}{57}},
  \bibinfo{pages}{835 } (\bibinfo{year}{2019}).

\bibitem[{\citenamefont{Vaqueiro et~al.}(2017)\citenamefont{Vaqueiro, Guelou,
  Kaltzoglou, Smith, Barbier, Guilmeau, and Powell}}]{vaqueiro2017influence}
\bibinfo{author}{\bibfnamefont{P.}~\bibnamefont{Vaqueiro}},
  \bibinfo{author}{\bibfnamefont{G.}~\bibnamefont{Guelou}},
  \bibinfo{author}{\bibfnamefont{A.}~\bibnamefont{Kaltzoglou}},
  \bibinfo{author}{\bibfnamefont{R.~I.} \bibnamefont{Smith}},
  \bibinfo{author}{\bibfnamefont{T.}~\bibnamefont{Barbier}},
  \bibinfo{author}{\bibfnamefont{E.}~\bibnamefont{Guilmeau}}, \bibnamefont{and}
  \bibinfo{author}{\bibfnamefont{A.~V.} \bibnamefont{Powell}},
  \bibinfo{journal}{Chemistry of Materials} \textbf{\bibinfo{volume}{29}},
  \bibinfo{pages}{4080} (\bibinfo{year}{2017}).

\bibitem[{\citenamefont{Tatsuka and Morimoto}(1977)}]{tatsuka1977tetrahedrite}
\bibinfo{author}{\bibfnamefont{K.}~\bibnamefont{Tatsuka}} \bibnamefont{and}
  \bibinfo{author}{\bibfnamefont{N.}~\bibnamefont{Morimoto}},
  \bibinfo{journal}{Economic Geology} \textbf{\bibinfo{volume}{72}},
  \bibinfo{pages}{258} (\bibinfo{year}{1977}).

\bibitem[{\citenamefont{Johnson and Jeanloz}(1983)}]{johnson1983brillouin}
\bibinfo{author}{\bibfnamefont{M.~L.} \bibnamefont{Johnson}} \bibnamefont{and}
  \bibinfo{author}{\bibfnamefont{R.}~\bibnamefont{Jeanloz}},
  \bibinfo{journal}{American Mineralogist} \textbf{\bibinfo{volume}{68}},
  \bibinfo{pages}{220} (\bibinfo{year}{1983}).

\bibitem[{\citenamefont{Suekuni et~al.}(2012)\citenamefont{Suekuni, Tsuruta,
  Ariga, and Koyano}}]{Suekuni_2012}
\bibinfo{author}{\bibfnamefont{K.}~\bibnamefont{Suekuni}},
  \bibinfo{author}{\bibfnamefont{K.}~\bibnamefont{Tsuruta}},
  \bibinfo{author}{\bibfnamefont{T.}~\bibnamefont{Ariga}}, \bibnamefont{and}
  \bibinfo{author}{\bibfnamefont{M.}~\bibnamefont{Koyano}},
  \bibinfo{journal}{Applied Physics Express} \textbf{\bibinfo{volume}{5}},
  \bibinfo{pages}{051201} (\bibinfo{year}{2012}).

\bibitem[{\citenamefont{Di~Benedetto et~al.}(2005)\citenamefont{Di~Benedetto,
  Bernardini, Cipriani, Emiliani, Gatteschi, and Romanelli}}]{DiBenedetto2005}
\bibinfo{author}{\bibfnamefont{F.}~\bibnamefont{Di~Benedetto}},
  \bibinfo{author}{\bibfnamefont{G.~P.} \bibnamefont{Bernardini}},
  \bibinfo{author}{\bibfnamefont{C.}~\bibnamefont{Cipriani}},
  \bibinfo{author}{\bibfnamefont{C.}~\bibnamefont{Emiliani}},
  \bibinfo{author}{\bibfnamefont{D.}~\bibnamefont{Gatteschi}},
  \bibnamefont{and}
  \bibinfo{author}{\bibfnamefont{M.}~\bibnamefont{Romanelli}},
  \bibinfo{journal}{Physics and Chemistry of Minerals}
  \textbf{\bibinfo{volume}{32}}, \bibinfo{pages}{155} (\bibinfo{year}{2005}).

\bibitem[{\citenamefont{Yan et~al.}(2018)\citenamefont{Yan, Wu, Wang, Lu, and
  Zhou}}]{YAN2018127}
\bibinfo{author}{\bibfnamefont{Y.}~\bibnamefont{Yan}},
  \bibinfo{author}{\bibfnamefont{H.}~\bibnamefont{Wu}},
  \bibinfo{author}{\bibfnamefont{G.}~\bibnamefont{Wang}},
  \bibinfo{author}{\bibfnamefont{X.}~\bibnamefont{Lu}}, \bibnamefont{and}
  \bibinfo{author}{\bibfnamefont{X.}~\bibnamefont{Zhou}},
  \bibinfo{journal}{Energy Storage Materials} \textbf{\bibinfo{volume}{13}},
  \bibinfo{pages}{127 } (\bibinfo{year}{2018}).

\bibitem[{\citenamefont{Bouyrie et~al.}(2015)\citenamefont{Bouyrie, Candolfi,
  Dauscher, Malaman, and Lenoir}}]{bouyrie2015exsolution}
\bibinfo{author}{\bibfnamefont{Y.}~\bibnamefont{Bouyrie}},
  \bibinfo{author}{\bibfnamefont{C.}~\bibnamefont{Candolfi}},
  \bibinfo{author}{\bibfnamefont{A.}~\bibnamefont{Dauscher}},
  \bibinfo{author}{\bibfnamefont{B.}~\bibnamefont{Malaman}}, \bibnamefont{and}
  \bibinfo{author}{\bibfnamefont{B.}~\bibnamefont{Lenoir}},
  \bibinfo{journal}{Chemistry of Materials} \textbf{\bibinfo{volume}{27}},
  \bibinfo{pages}{8354} (\bibinfo{year}{2015}).

\bibitem[{\citenamefont{Jana and Biswas}(2018)}]{Jana2018}
\bibinfo{author}{\bibfnamefont{M.~K.} \bibnamefont{Jana}} \bibnamefont{and}
  \bibinfo{author}{\bibfnamefont{K.}~\bibnamefont{Biswas}},
  \bibinfo{journal}{ACS Energy Letters} \textbf{\bibinfo{volume}{3}},
  \bibinfo{pages}{1315} (\bibinfo{year}{2018}).

\bibitem[{\citenamefont{Munjal and Khare}(2019)}]{Munjal_2019}
\bibinfo{author}{\bibfnamefont{S.}~\bibnamefont{Munjal}} \bibnamefont{and}
  \bibinfo{author}{\bibfnamefont{N.}~\bibnamefont{Khare}},
  \bibinfo{journal}{Journal of Physics D: Applied Physics}
  \textbf{\bibinfo{volume}{52}}, \bibinfo{pages}{433002}
  (\bibinfo{year}{2019}).

\bibitem[{\citenamefont{Brinkmann}(1992)}]{brinkmann1992nmr}
\bibinfo{author}{\bibfnamefont{D.}~\bibnamefont{Brinkmann}},
  \bibinfo{journal}{Progress in Nuclear Magnetic Resonance Spectroscopy}
  \textbf{\bibinfo{volume}{24}}, \bibinfo{pages}{527} (\bibinfo{year}{1992}).

\bibitem[{\citenamefont{Hull}(2004)}]{hull2004superionics}
\bibinfo{author}{\bibfnamefont{S.}~\bibnamefont{Hull}},
  \bibinfo{journal}{Reports on Progress in Physics}
  \textbf{\bibinfo{volume}{67}}, \bibinfo{pages}{1233} (\bibinfo{year}{2004}).

\bibitem[{\citenamefont{Tang et~al.}(1999)\citenamefont{Tang, Geyer, Busch,
  Johnson, and Wu}}]{tang1999diffusion}
\bibinfo{author}{\bibfnamefont{X.-P.} \bibnamefont{Tang}},
  \bibinfo{author}{\bibfnamefont{U.}~\bibnamefont{Geyer}},
  \bibinfo{author}{\bibfnamefont{R.}~\bibnamefont{Busch}},
  \bibinfo{author}{\bibfnamefont{W.~L.} \bibnamefont{Johnson}},
  \bibnamefont{and} \bibinfo{author}{\bibfnamefont{Y.}~\bibnamefont{Wu}},
  \bibinfo{journal}{Nature} \textbf{\bibinfo{volume}{402}},
  \bibinfo{pages}{160} (\bibinfo{year}{1999}).

\bibitem[{\citenamefont{Toby and Von~Dreele}(2013)}]{toby2013gsas}
\bibinfo{author}{\bibfnamefont{B.~H.} \bibnamefont{Toby}} \bibnamefont{and}
  \bibinfo{author}{\bibfnamefont{R.~B.} \bibnamefont{Von~Dreele}},
  \bibinfo{journal}{Journal of Applied Crystallography}
  \textbf{\bibinfo{volume}{46}}, \bibinfo{pages}{544} (\bibinfo{year}{2013}).

\bibitem[{\citenamefont{Carter et~al.}(1976)\citenamefont{Carter, Bennett, and
  Kahan}}]{19761}
\bibinfo{author}{\bibfnamefont{G.}~\bibnamefont{Carter}},
  \bibinfo{author}{\bibfnamefont{L.}~\bibnamefont{Bennett}}, \bibnamefont{and}
  \bibinfo{author}{\bibfnamefont{D.}~\bibnamefont{Kahan}},
  \bibinfo{journal}{Progress in Materials Science}
  \textbf{\bibinfo{volume}{20}}, \bibinfo{pages}{1 } (\bibinfo{year}{1976}).

\bibitem[{\citenamefont{Harris and Becker}(2002)}]{harris2002nmr}
\bibinfo{author}{\bibfnamefont{R.~K.} \bibnamefont{Harris}} \bibnamefont{and}
  \bibinfo{author}{\bibfnamefont{E.~D.} \bibnamefont{Becker}},
  \bibinfo{journal}{Journal of Magnetic Resonance}
  \textbf{\bibinfo{volume}{2}}, \bibinfo{pages}{323} (\bibinfo{year}{2002}).

\bibitem[{\citenamefont{Ghassemi et~al.}(2018)\citenamefont{Ghassemi, Lu, Tian,
  Conant, Yan, Zhou, and Ross~Jr}}]{ghassemi2018structure}
\bibinfo{author}{\bibfnamefont{N.}~\bibnamefont{Ghassemi}},
  \bibinfo{author}{\bibfnamefont{X.}~\bibnamefont{Lu}},
  \bibinfo{author}{\bibfnamefont{Y.}~\bibnamefont{Tian}},
  \bibinfo{author}{\bibfnamefont{E.}~\bibnamefont{Conant}},
  \bibinfo{author}{\bibfnamefont{Y.}~\bibnamefont{Yan}},
  \bibinfo{author}{\bibfnamefont{X.}~\bibnamefont{Zhou}}, \bibnamefont{and}
  \bibinfo{author}{\bibfnamefont{J.~H.} \bibnamefont{Ross~Jr}},
  \bibinfo{journal}{ACS Applied Materials \& Interfaces}
  \textbf{\bibinfo{volume}{10}}, \bibinfo{pages}{36010} (\bibinfo{year}{2018}).

\bibitem[{\citenamefont{Bastow et~al.}(2015)\citenamefont{Bastow, Lehmann-Horn,
  and Miljak}}]{bastow2015121}
\bibinfo{author}{\bibfnamefont{T.}~\bibnamefont{Bastow}},
  \bibinfo{author}{\bibfnamefont{J.}~\bibnamefont{Lehmann-Horn}},
  \bibnamefont{and} \bibinfo{author}{\bibfnamefont{D.}~\bibnamefont{Miljak}},
  \bibinfo{journal}{Solid State Nuclear Magnetic Resonance}
  \textbf{\bibinfo{volume}{71}}, \bibinfo{pages}{55} (\bibinfo{year}{2015}).

\bibitem[{\citenamefont{Matsui et~al.}(2019)\citenamefont{Matsui, Matsuno,
  Kotegawa, Tou, Suekuni, Hasegawa, Tanaka, and Takabatake}}]{matsui2019first}
\bibinfo{author}{\bibfnamefont{T.}~\bibnamefont{Matsui}},
  \bibinfo{author}{\bibfnamefont{H.}~\bibnamefont{Matsuno}},
  \bibinfo{author}{\bibfnamefont{H.}~\bibnamefont{Kotegawa}},
  \bibinfo{author}{\bibfnamefont{H.}~\bibnamefont{Tou}},
  \bibinfo{author}{\bibfnamefont{K.}~\bibnamefont{Suekuni}},
  \bibinfo{author}{\bibfnamefont{T.}~\bibnamefont{Hasegawa}},
  \bibinfo{author}{\bibfnamefont{H.~I.} \bibnamefont{Tanaka}},
  \bibnamefont{and}
  \bibinfo{author}{\bibfnamefont{T.}~\bibnamefont{Takabatake}},
  \bibinfo{journal}{Journal of the Physical Society of Japan}
  \textbf{\bibinfo{volume}{88}}, \bibinfo{pages}{054710}
  (\bibinfo{year}{2019}).

\bibitem[{\citenamefont{Kitagawa et~al.}(2015)\citenamefont{Kitagawa, Sekiya,
  Araki, Kobayashi, Ishida, Kambe, Kimura, Nishimoto, Kudo, and
  Nohara}}]{kitagawa2015suppression}
\bibinfo{author}{\bibfnamefont{S.}~\bibnamefont{Kitagawa}},
  \bibinfo{author}{\bibfnamefont{T.}~\bibnamefont{Sekiya}},
  \bibinfo{author}{\bibfnamefont{S.}~\bibnamefont{Araki}},
  \bibinfo{author}{\bibfnamefont{T.~C.} \bibnamefont{Kobayashi}},
  \bibinfo{author}{\bibfnamefont{K.}~\bibnamefont{Ishida}},
  \bibinfo{author}{\bibfnamefont{T.}~\bibnamefont{Kambe}},
  \bibinfo{author}{\bibfnamefont{T.}~\bibnamefont{Kimura}},
  \bibinfo{author}{\bibfnamefont{N.}~\bibnamefont{Nishimoto}},
  \bibinfo{author}{\bibfnamefont{K.}~\bibnamefont{Kudo}}, \bibnamefont{and}
  \bibinfo{author}{\bibfnamefont{M.}~\bibnamefont{Nohara}},
  \bibinfo{journal}{Journal of the Physical Society of Japan}
  \textbf{\bibinfo{volume}{84}}, \bibinfo{pages}{093701}
  (\bibinfo{year}{2015}).

\bibitem[{\citenamefont{Abragam}(1981)}]{Abragam}
\bibinfo{author}{\bibfnamefont{A.}~\bibnamefont{Abragam}},
  \emph{\bibinfo{title}{The Principles of Nuclear Magnetism}}
  (\bibinfo{publisher}{Oxford science publications}, \bibinfo{year}{1981}).

\bibitem[{\citenamefont{Boyce et~al.}(1979)\citenamefont{Boyce, Mikkelsen, and
  Huberman}}]{BOYCE1979507}
\bibinfo{author}{\bibfnamefont{J.}~\bibnamefont{Boyce}},
  \bibinfo{author}{\bibfnamefont{J.}~\bibnamefont{Mikkelsen}},
  \bibnamefont{and} \bibinfo{author}{\bibfnamefont{B.}~\bibnamefont{Huberman}},
  \bibinfo{journal}{Solid State Communications} \textbf{\bibinfo{volume}{29}},
  \bibinfo{pages}{507 } (\bibinfo{year}{1979}).

\bibitem[{\citenamefont{Eguchi et~al.}(1981)\citenamefont{Eguchi, Marinos,
  Jonas, Silbernagel, and Thompson}}]{eguchi1981high}
\bibinfo{author}{\bibfnamefont{T.}~\bibnamefont{Eguchi}},
  \bibinfo{author}{\bibfnamefont{C.}~\bibnamefont{Marinos}},
  \bibinfo{author}{\bibfnamefont{J.}~\bibnamefont{Jonas}},
  \bibinfo{author}{\bibfnamefont{B.}~\bibnamefont{Silbernagel}},
  \bibnamefont{and} \bibinfo{author}{\bibfnamefont{A.}~\bibnamefont{Thompson}},
  \bibinfo{journal}{Solid State Communications} \textbf{\bibinfo{volume}{38}},
  \bibinfo{pages}{919} (\bibinfo{year}{1981}).

\bibitem[{\citenamefont{Makovicky and Skinner}(1979)}]{makovicky1979studies}
\bibinfo{author}{\bibfnamefont{E.}~\bibnamefont{Makovicky}} \bibnamefont{and}
  \bibinfo{author}{\bibfnamefont{B.~J.} \bibnamefont{Skinner}},
  \bibinfo{journal}{The Canadian Mineralogist} \textbf{\bibinfo{volume}{17}},
  \bibinfo{pages}{619} (\bibinfo{year}{1979}).

\bibitem[{\citenamefont{Bullett and Dawson}(1986)}]{Bullett_1986}
\bibinfo{author}{\bibfnamefont{D.~W.} \bibnamefont{Bullett}} \bibnamefont{and}
  \bibinfo{author}{\bibfnamefont{W.~G.} \bibnamefont{Dawson}},
  \bibinfo{journal}{Journal of Physics C: Solid State Physics}
  \textbf{\bibinfo{volume}{19}}, \bibinfo{pages}{5837} (\bibinfo{year}{1986}).

\bibitem[{\citenamefont{Chen and Mak}(1993)}]{chen1993metal}
\bibinfo{author}{\bibfnamefont{X.-M.} \bibnamefont{Chen}} \bibnamefont{and}
  \bibinfo{author}{\bibfnamefont{T.~C.} \bibnamefont{Mak}},
  \bibinfo{journal}{Structural Chemistry} \textbf{\bibinfo{volume}{4}},
  \bibinfo{pages}{247} (\bibinfo{year}{1993}).

\bibitem[{\citenamefont{Mehrotra and Hoffmann}(1978)}]{mehrotra1978copper}
\bibinfo{author}{\bibfnamefont{P.~K.} \bibnamefont{Mehrotra}} \bibnamefont{and}
  \bibinfo{author}{\bibfnamefont{R.}~\bibnamefont{Hoffmann}},
  \bibinfo{journal}{Inorganic Chemistry} \textbf{\bibinfo{volume}{17}},
  \bibinfo{pages}{2187} (\bibinfo{year}{1978}).

\bibitem[{\citenamefont{Sirusi and Ross}(2017)}]{SIRUSI2017137}
\bibinfo{author}{\bibfnamefont{A.~A.} \bibnamefont{Sirusi}} \bibnamefont{and}
  \bibinfo{author}{\bibfnamefont{J.~H.} \bibnamefont{Ross}},
  \bibinfo{journal}{Annual Reports on NMR Spectroscopy}
  \textbf{\bibinfo{volume}{92}}, \bibinfo{pages}{137 } (\bibinfo{year}{2017}).

\bibitem[{\citenamefont{Lu et~al.}(2015)\citenamefont{Lu, Morelli, Xia, and
  Ozolins}}]{lu2015increasing}
\bibinfo{author}{\bibfnamefont{X.}~\bibnamefont{Lu}},
  \bibinfo{author}{\bibfnamefont{D.~T.} \bibnamefont{Morelli}},
  \bibinfo{author}{\bibfnamefont{Y.}~\bibnamefont{Xia}}, \bibnamefont{and}
  \bibinfo{author}{\bibfnamefont{V.}~\bibnamefont{Ozolins}},
  \bibinfo{journal}{Chemistry of Materials} \textbf{\bibinfo{volume}{27}},
  \bibinfo{pages}{408} (\bibinfo{year}{2015}).

\bibitem[{\citenamefont{Zha et~al.}(1996)\citenamefont{Zha, Barzykin, and
  Pines}}]{PhysRevB.54.7561}
\bibinfo{author}{\bibfnamefont{Y.}~\bibnamefont{Zha}},
  \bibinfo{author}{\bibfnamefont{V.}~\bibnamefont{Barzykin}}, \bibnamefont{and}
  \bibinfo{author}{\bibfnamefont{D.}~\bibnamefont{Pines}},
  \bibinfo{journal}{Phys. Rev. B} \textbf{\bibinfo{volume}{54}},
  \bibinfo{pages}{7561} (\bibinfo{year}{1996}).

\bibitem[{\citenamefont{Wang et~al.}(2015)\citenamefont{Wang, Richards, Ong,
  Miara, Kim, Mo, and Ceder}}]{wang2015design}
\bibinfo{author}{\bibfnamefont{Y.}~\bibnamefont{Wang}},
  \bibinfo{author}{\bibfnamefont{W.~D.} \bibnamefont{Richards}},
  \bibinfo{author}{\bibfnamefont{S.~P.} \bibnamefont{Ong}},
  \bibinfo{author}{\bibfnamefont{L.~J.} \bibnamefont{Miara}},
  \bibinfo{author}{\bibfnamefont{J.~C.} \bibnamefont{Kim}},
  \bibinfo{author}{\bibfnamefont{Y.}~\bibnamefont{Mo}}, \bibnamefont{and}
  \bibinfo{author}{\bibfnamefont{G.}~\bibnamefont{Ceder}},
  \bibinfo{journal}{Nature Materials} \textbf{\bibinfo{volume}{14}},
  \bibinfo{pages}{1026} (\bibinfo{year}{2015}).

\bibitem[{\citenamefont{Sirusi et~al.}(2015)\citenamefont{Sirusi, Ballikaya,
  Uher, and Ross}}]{doi:10.1021/acs.jpcc.5b06079}
\bibinfo{author}{\bibfnamefont{A.~A.} \bibnamefont{Sirusi}},
  \bibinfo{author}{\bibfnamefont{S.}~\bibnamefont{Ballikaya}},
  \bibinfo{author}{\bibfnamefont{C.}~\bibnamefont{Uher}}, \bibnamefont{and}
  \bibinfo{author}{\bibfnamefont{J.~H.} \bibnamefont{Ross}},
  \bibinfo{journal}{The Journal of Physical Chemistry C}
  \textbf{\bibinfo{volume}{119}}, \bibinfo{pages}{20293}
  (\bibinfo{year}{2015}).

\bibitem[{\citenamefont{Sirusi et~al.}(2016)\citenamefont{Sirusi, Ballikaya,
  Chen, Uher, and Ross~Jr}}]{sirusi2016band}
\bibinfo{author}{\bibfnamefont{A.~A.} \bibnamefont{Sirusi}},
  \bibinfo{author}{\bibfnamefont{S.}~\bibnamefont{Ballikaya}},
  \bibinfo{author}{\bibfnamefont{J.-H.} \bibnamefont{Chen}},
  \bibinfo{author}{\bibfnamefont{C.}~\bibnamefont{Uher}}, \bibnamefont{and}
  \bibinfo{author}{\bibfnamefont{J.~H.} \bibnamefont{Ross~Jr}},
  \bibinfo{journal}{The Journal of Physical Chemistry C}
  \textbf{\bibinfo{volume}{120}}, \bibinfo{pages}{14549}
  (\bibinfo{year}{2016}).

\bibitem[{\citenamefont{Yesinowski et~al.}(2010)\citenamefont{Yesinowski,
  Ladouceur, Purdy, and Miller}}]{yesinowski2010electrical}
\bibinfo{author}{\bibfnamefont{J.~P.} \bibnamefont{Yesinowski}},
  \bibinfo{author}{\bibfnamefont{H.~D.} \bibnamefont{Ladouceur}},
  \bibinfo{author}{\bibfnamefont{A.~P.} \bibnamefont{Purdy}}, \bibnamefont{and}
  \bibinfo{author}{\bibfnamefont{J.~B.} \bibnamefont{Miller}},
  \bibinfo{journal}{The Journal of Chemical Physics}
  \textbf{\bibinfo{volume}{133}}, \bibinfo{pages}{234509}
  (\bibinfo{year}{2010}).

\bibitem[{\citenamefont{Blaha et~al.}(2001)\citenamefont{Blaha, Schwarz,
  Madsen, Kvasnicka, and Luitz}}]{blaha2001wien2k}
\bibinfo{author}{\bibfnamefont{P.}~\bibnamefont{Blaha}},
  \bibinfo{author}{\bibfnamefont{K.}~\bibnamefont{Schwarz}},
  \bibinfo{author}{\bibfnamefont{G.~K.} \bibnamefont{Madsen}},
  \bibinfo{author}{\bibfnamefont{D.}~\bibnamefont{Kvasnicka}},
  \bibnamefont{and} \bibinfo{author}{\bibfnamefont{J.}~\bibnamefont{Luitz}},
  \bibinfo{journal}{An augmented plane wave+ local orbitals program for
  calculating crystal properties}  (\bibinfo{year}{2001}).

\bibitem[{\citenamefont{Kresse and
  Furthm{\"u}ller}(1996)}]{kresse1996efficient}
\bibinfo{author}{\bibfnamefont{G.}~\bibnamefont{Kresse}} \bibnamefont{and}
  \bibinfo{author}{\bibfnamefont{J.}~\bibnamefont{Furthm{\"u}ller}},
  \bibinfo{journal}{Physical review B} \textbf{\bibinfo{volume}{54}},
  \bibinfo{pages}{11169} (\bibinfo{year}{1996}).

\bibitem[{\citenamefont{Pfitzner and Reiser}(2002)}]{pfitzner2002refinement}
\bibinfo{author}{\bibfnamefont{A.}~\bibnamefont{Pfitzner}} \bibnamefont{and}
  \bibinfo{author}{\bibfnamefont{S.}~\bibnamefont{Reiser}},
  \bibinfo{journal}{Zeitschrift f{\"u}r Kristallographie-Crystalline Materials}
  \textbf{\bibinfo{volume}{217}}, \bibinfo{pages}{51} (\bibinfo{year}{2002}).

\bibitem[{\citenamefont{Wang et~al.}(2011)\citenamefont{Wang, Lin, Das, Hasan,
  and Bansil}}]{wang2011topological}
\bibinfo{author}{\bibfnamefont{Y.}~\bibnamefont{Wang}},
  \bibinfo{author}{\bibfnamefont{H.}~\bibnamefont{Lin}},
  \bibinfo{author}{\bibfnamefont{T.}~\bibnamefont{Das}},
  \bibinfo{author}{\bibfnamefont{M.}~\bibnamefont{Hasan}}, \bibnamefont{and}
  \bibinfo{author}{\bibfnamefont{A.}~\bibnamefont{Bansil}},
  \bibinfo{journal}{New Journal of Physics} \textbf{\bibinfo{volume}{13}},
  \bibinfo{pages}{085017} (\bibinfo{year}{2011}).

\bibitem[{\citenamefont{Crespo}(2016)}]{crespo2016microscopic}
\bibinfo{author}{\bibfnamefont{C.~T.} \bibnamefont{Crespo}},
  \bibinfo{journal}{The Journal of Physical Chemistry C}
  \textbf{\bibinfo{volume}{120}}, \bibinfo{pages}{7959} (\bibinfo{year}{2016}).

\end{thebibliography}

\vspace{6cm}

\newcommand{\hbAppendixPrefix}{S}
\renewcommand{\thefigure}{\hbAppendixPrefix\arabic{figure}}
\setcounter{figure}{0}

\section*{Supporting Information}

\begin{figure}[h]
\includegraphics[width=8cm]{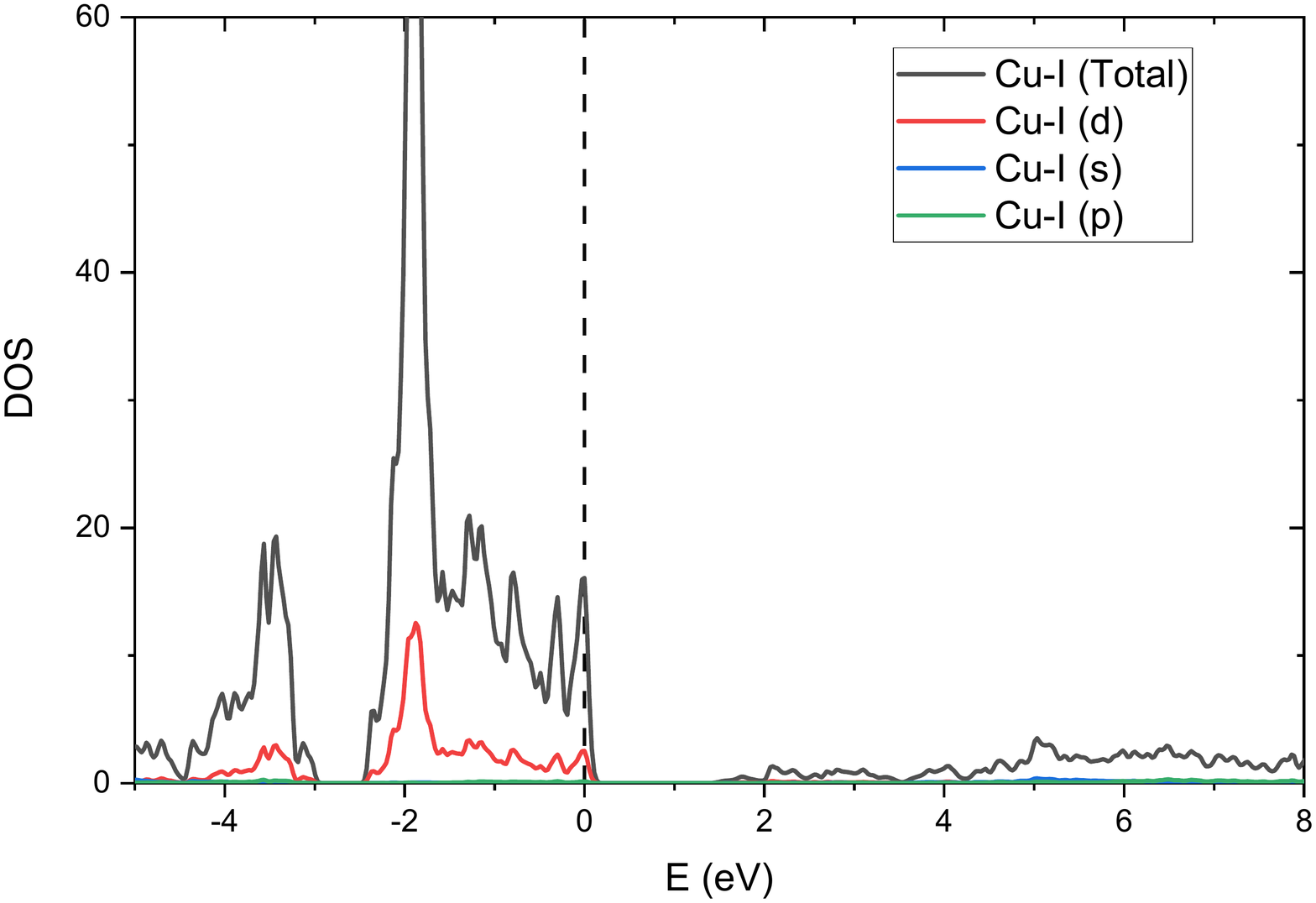}
\centering
\caption{ Partial density of states of Cu-I for \ch{Cu12Sb4S13}.}
\label{fig: S1}
\end{figure}

\begin{figure}[h]
\includegraphics[width=8cm]{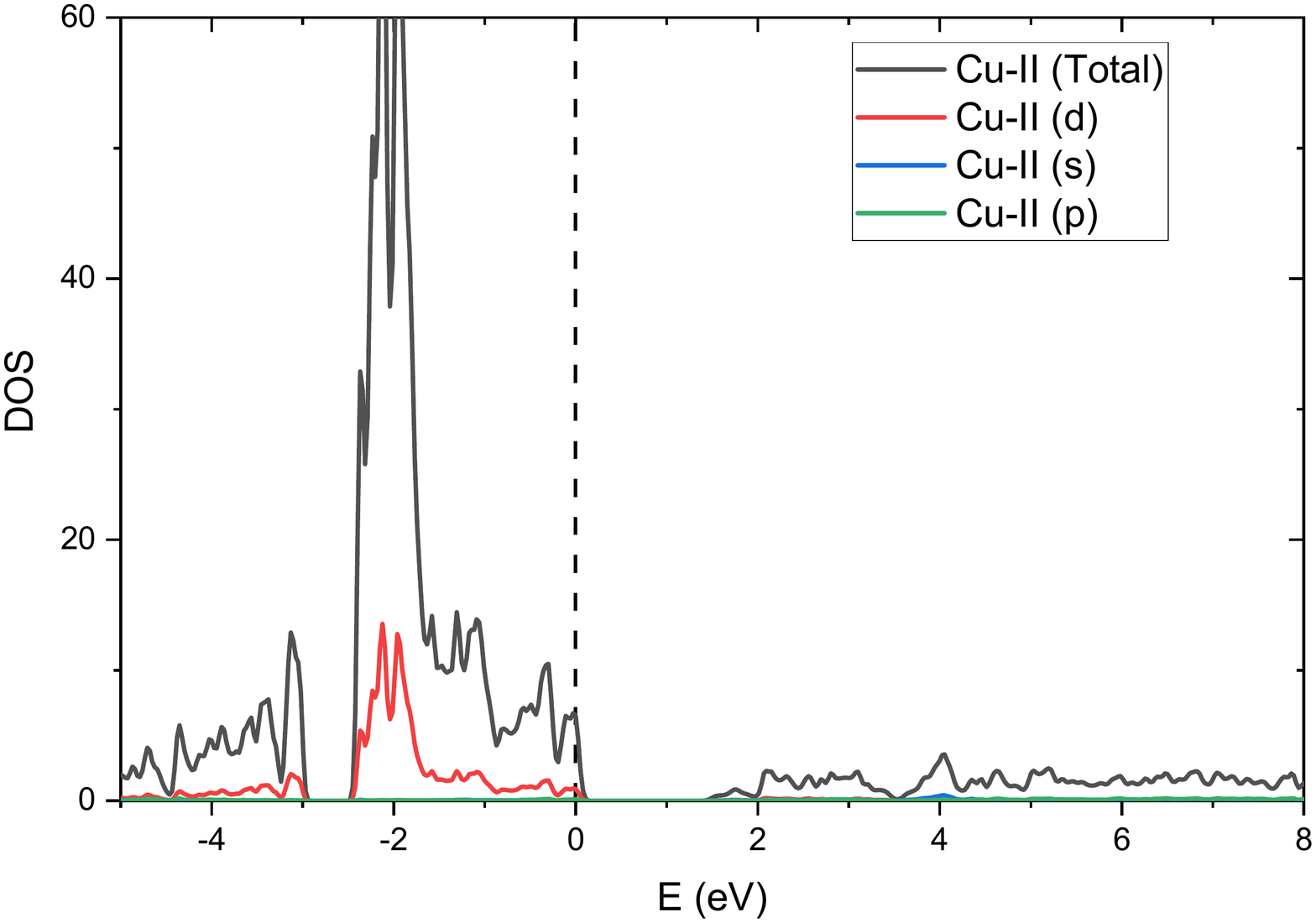}
\centering
\caption{ Partial density of states of Cu-II for \ch{Cu12Sb4S13}.}
\label{fig: S2}
\end{figure}

\begin{figure}[h]
\includegraphics[width=8cm]{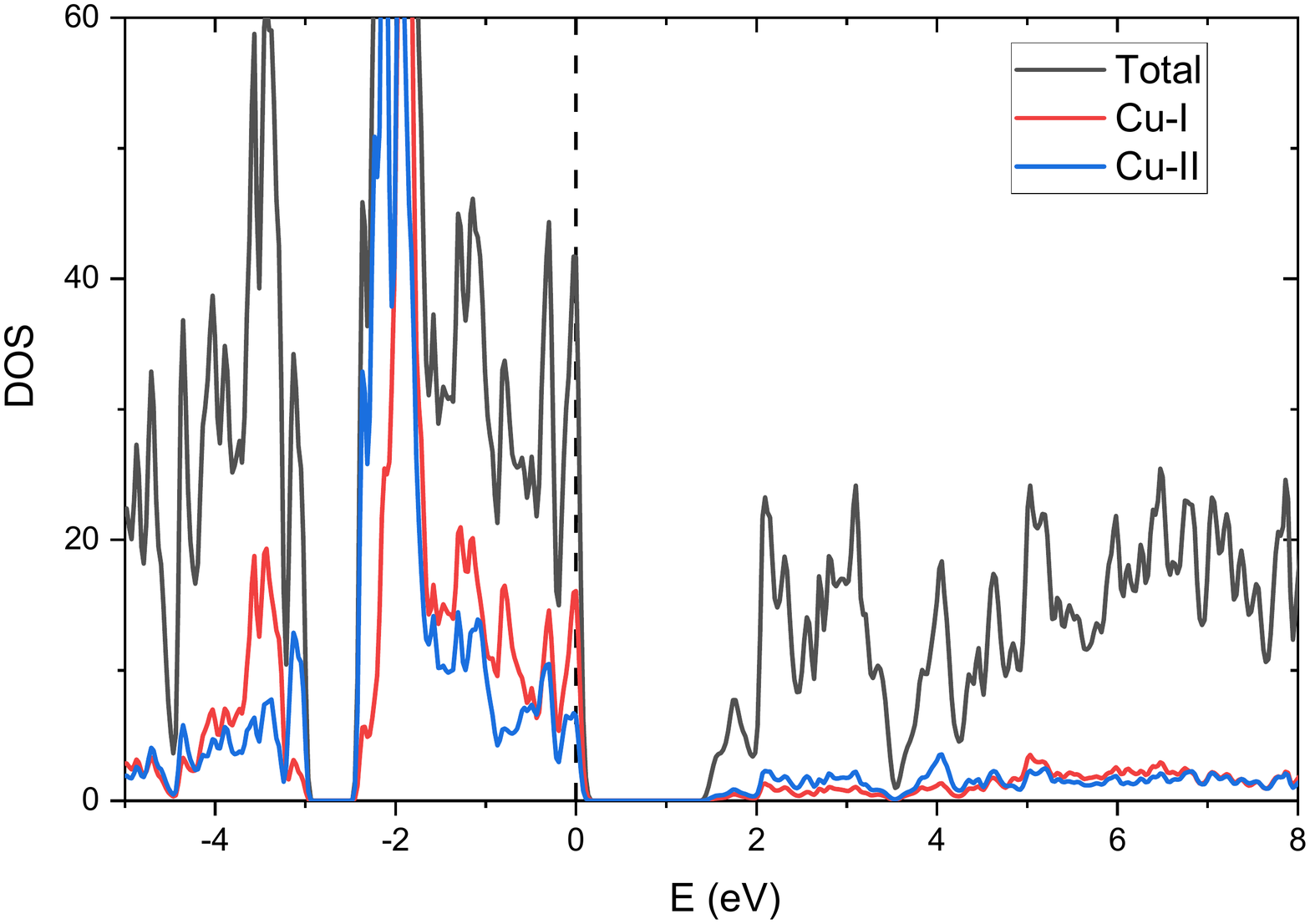}
\centering
\caption{Partial density of states for Cu-I and Cu-II and total density of states for \ch{Cu12Sb4S13}.}
\label{fig: S3}
\end{figure}

Density function theory (DFT) calculations were conducted using the WIEN2k package \cite{blaha2001wien2k}. In this package, a (linearized) augmented plane wave plus local orbitals method is implemented \cite{kresse1996efficient}. For these calculations, a cubic unit cell of with $58$ atoms in space group $217$ with a lattice constant of $a = \SI{10.3208}{\angstrom}$ was considered. We used the atomic positions obtained in the XRD refinements for the \ch{Cu12Sb4S13} phase. We took separation energy between core and valence states as $RK_{max}$= $6$ Ry, the plane-wave expansion cutoff $G_{max}= 12$ $Bohr^{-1}$, and $12 \times 12 \times 12$ k-points and adopted the exchange correlation functional introduced by Perdew, Burke, and Ernzerhof (PBE) \cite{kresse1996efficient}. The calculation was run without spin-orbit coupling or spin polarization. Density of states results are shown in Figure \ref{fig: S1}, for the Cu-I site, and in Figure \ref{fig: S2} for the Cu-II site. Also, Figure \ref{fig: S3} shows a comparison of the full density of states and the Cu partial densities of states (for all orbital symmetries) for the two sites. The full density of states is comparable to what has been reported previously \cite{lu2013high}.

\begin{figure}[h]
\includegraphics[width=12cm]{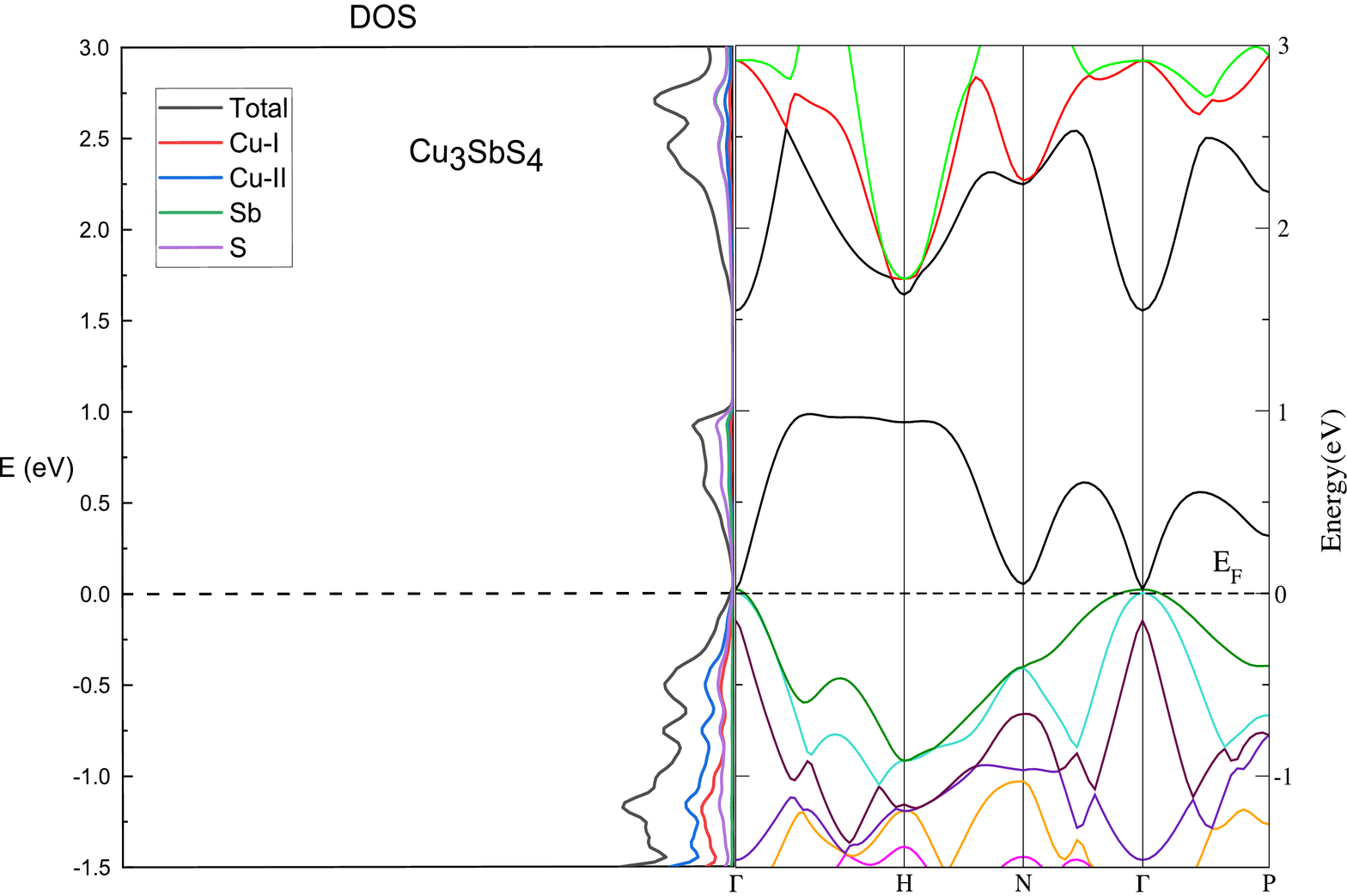}
\centering
\caption{Density of states and energy-band diagram for Cu3SbS4 along high symmetry lines in the BZ.}
\label{fig: S4}
\end{figure}

For \ch{Cu3SbS4}, we used the structure reported by Pfitzer \cite{pfitzner2002refinement} which includes a tetragonal unit cell of $8$ atoms in $121(I\overline{4}2)$ space group with $a = \SI{5.391}{\angstrom}$ and $c = \SI{10.764}{\angstrom}$. The calculation initialized with $RK_{max}= 6$, the plane-wave expansion cutoff of $G_{max}= 12$, $Bohr^{-1}$ and $10\times 10\times10$ k-points considered with PBE (Perdew, Burke, and Ernzerhof) exchange correlation function, without spin polarization or the spin orbit coupling. Our results are in good agreement with recently reported calculations \cite{wang2011topological}. However, note that it was recently shown \cite{crespo2016microscopic} that using more advanced functionals yields a nonzero semiconducting gap for this material. However, this is unlikely to have a large effect on calculated electric field gradients (EFGs), which depends only on the filled states within the valence band. The EFG results for the two Cu sites in this material, Cu-I and Cu-II, are $2.85318 \times 10^{21}$ ($V/m^2$) and $-1.46966 \times 10^{21}$ ($V/m^2$) respectively, quoted as the largest-magnitude principal values of the EFG tensors, which are symmetry-constrained to be axial in both cases. Converting to the standard quadrupole parameter $\nu_Q= \frac{3eQV_{zz}}{2I(2I-1)h}$, based on the standard \ce{^{63}Cu} quadrupole moment \cite{harris2002nmr}, we obtain $\nu_Q = 7.59$ MHz and $\nu_Q= 3.85$ MHz respectively.

\end{document}